\documentclass{aa}
\usepackage{graphicx}
\usepackage{grffile}
\usepackage{tablefootnote}
\usepackage{txfonts}
\usepackage{xcolor}
\usepackage{color}
\usepackage{amsmath}    
\usepackage{txfonts}
\usepackage{hyperref}
\usepackage{multirow}
\usepackage{notes2bib}
\usepackage{stfloats}
\usepackage{float}
\usepackage[normalem]{ulem}
\usepackage{comment}

\restylefloat{table}

\hypersetup{colorlinks=true,linkcolor=[rgb]{1.,0.2,0.2},citecolor=[rgb]{0.1,0.4,1.},filecolor=[rgb]{0.7,0.2,0.2},urlcolor=[rgb]{0.0,0.2,1.}}
\definecolor{darkgreen}{rgb}{0.09, 0.45, 0.27}
\definecolor{amber(sae/ece)}{rgb}{1.0, 0.49, 0.0}

\newcommand{\Reff}{$\mathrm{R}_{\mathrm{e}\,}$}
\newcommand{\ppxf}{\textsc{pPXF}}
\newcommand{\mue}{$\langle\mu_{\mathrm {e}}\rangle \,$}
\newcommand{\kms}{km s$^{-1}$}
\newcommand{\Msun}{M$_{\odot}\,$}
\newcommand{\Mstar}{M$_{\star}\,$}

\def\INSPIRE{\mbox{{\tt INSPIRE}}}

\begin{document} 

%%%%%  TITLE %%%%%
   \title{INSPIRE: INvestigating Stellar Population In RElics}
   \subtitle{I. Survey presentation and pilot study}

%%%%% AUTHORS %%%%%
   \author{C.~Spiniello\inst{\ref{oxf},\ref{inaf_naples}}\fnmsep\thanks{Corresponding author: {\tt chiara.spiniello@physics.ox.ac.uk}.}  
   %GROUP 1 
   \and C.~Tortora\inst{\ref{inaf_naples},\ref{inaf_arcetri}}
   \and 
   G.~D'Ago\inst{\ref{puc}}
   \and L.~Coccato\inst{\ref{eso}}
   %GROUP 2
   \and F.~La Barbera\inst{\ref{inaf_naples}}
   \and A.~Ferr\'e-Mateu\inst{\ref{iccub}} 
   \and \\
   N.~R.~Napolitano\inst{\ref{SunYat}, \ref{inaf_naples}}
   \and M.~Spavone\inst{\ref{inaf_naples}}
   \and D.~Scognamiglio\inst{\ref{bonn}} 
      %GROUP 3 
   \and M.~Arnaboldi\inst{\ref{eso}}
   \and A.~Gallazzi\inst{\ref{inaf_arcetri}}
   \and L.~Hunt\inst{\ref{inaf_arcetri}}
   \and \\
   S.~Moehler\inst{\ref{eso}}
   \and M.~Radovich\inst{\ref{inaf_padova}}
   \and S. Zibetti\inst{\ref{inaf_arcetri}}
   }

   \institute{Sub-Dep. of Astrophysics, Dep. of Physics, University of Oxford, Denys Wilkinson Building, Keble Road, Oxford OX1 3RH, UK\label{oxf} 
    \and
    INAF -- Osservatorio Astronomico di Capodimonte, Via Moiariello  16, 80131, Naples, Italy\label{inaf_naples}
    \and
    INAF -- Osservatorio Astronomico di Arcetri, Largo Enrico Fermi 5, 50125, Firenze, Italy\label{inaf_arcetri}
    \and
    Institute of Astrophysics, Pontificia Universidad Cat\'olica de Chile, Av. Vicu\~na Mackenna 4860, 7820436 Macul, Santiago, Chile\label{puc}
    \and 
    European Southern Observatory,  Karl-Schwarzschild-Stra\ss{}e 2, 85748, Garching, Germany\label{eso} 
    \and
    Institut de Ciencies del Cosmos (ICCUB), Universitat de Barcelona (IEEC-UB), E02028 Barcelona, Spain\label{iccub} 
    %\and
    %Centre for Astrophysics \& Supercomputing, Swinburne University, Hawthorn VIC 3122, Australia\label{swim}
    \and
    School of Physics and Astronomy, Sun Yat-Sen University, 2 Daxue Road, Tangjia, Zhuhai, Guangdong 519082, P.R. China\label{SunYat} 
    \and
    Argelander-Institut f\"{u}r Astronomie, Auf dem H\"{u}gel 71, D-53121 - Bonn, Germany\label{bonn}
    \and 
    INAF -- Osservatorio Astronomico di Padova, Vicolo Osservatorio 5, I-35122 Padova, Italy\label{inaf_padova}
    }

   \date{Received 19 November 2020}

  \abstract
  % context heading (optional)
  % {} leave it empty if necessary  
   {Massive elliptical galaxies are thought to form through a two-phase process. At early times ($z>2$), an intense and fast starburst forms blue and disk-dominated galaxies. After quenching, the remaining structures become red, compact, and massive (i.e. red nuggets).  
%   first an intense star formation burst, after quenching, creates red, compact and massive objects (âred nuggetsâ) at $z \gtrsim 2$. 
    Then, a time-extended second phase, which is dominated by mergers, causes structural evolution and size growth. Given the stochastic nature of mergers, a small fraction of red nuggets survive, without any interaction, massive and compact until today: these are relic galaxies. Since this fraction depends on the processes dominating the size growth, counting relics at low-z is a valuable way of disentangling between different galaxy evolution models.} 
  % methods heading (mandatory)
   {In this paper, we introduce the INvestigating Stellar Population In RElics (\INSPIRE) Project, which aims to spectroscopically confirm and fully characterise a large number of relics at $0.1<z<0.5$. We focus here on the first results based on a pilot study targeting three systems, representative of the whole sample.} 
  % results heading (mandatory)
   {For these three candidates, we extracted 1D optical spectra over an aperture of $r=0.40\arcsec$, which comprises $\sim$30\% of the galaxies' light, and we obtained the line-of-sight integrated stellar velocity and velocity dispersion. We also inferred the stellar [$\alpha$/Fe] abundance from line-index measurements and mass-weighted age and metallicity from full-spectral fitting with single stellar population models.} %\LEt{Please note that the specific steps you took for this particular research should be in the simple past: We used, We extrapolated, They determined, etc.  If you are speaking about universal truths, constants, or findings, the present simple should be used: We find, The solution to the equation is, etc. If the research is ongoing, then the present perfect can be used: We have used, We have extrapolated, They have found, etc.  See Sect. 4 of the Language Guide (https://www.aanda.org/for-authors/language-editing/1-introduction). Please do check your past/present tenses carefully throughout, as I'm not always able to accurately tell the difference, and so I risk altering your meaning when I do make changes, and leaving a mistake when I dont. Thank you.}}
   {Two galaxies have large integrated stellar velocity dispersion values ($\sigma_{\star} \sim 250$ \kms), confirming their massive nature. They are populated by stars with super-solar metallicity and [$\alpha$/Fe]. Both objects  have formed $\ge80\%$ of their stellar mass within a short ($\sim 0.5-1.0$ Gyrs) initial star formation episode occurred only $\sim 1$ Gyr after the Big Bang. The third galaxy has a more extended star formation history and a lower velocity dispersion. Thus we confirm two out of three candidates as relics. }
    %A degree of relics, already proposed in the local Universe, seems to be confirmed too, but more data are needed.}
   { This paper is the first step towards assembling the final \INSPIRE\, catalogue that will set stringent lower limits on the number density of relics at $z<0.5$, thus constituting a benchmark for cosmological simulations, and their predictions on number densities, sizes, masses, and dynamical characteristics of these objects.}

  %One galaxy, KiDS J0847+0112, is doubtless the most extreme case, as it assembled the totality of its mass during the first burst and then had no further sign of star formation after it. For the other system the situation is less extreme, as it shows minor secondary star formation episodes. Thus, 
   %J0847+0112 and J0224-3143 are the first two confirmed relics at intermediate redshift, as they had no further sign of star formation after the first burst. J0314-3215 assembled roughly 80\% of its stellar mass during the first episode, but then it experienced a second one $\sim5.5$ Gyrs after the BB. J0224-3143 had two major episodes. In conclusion, a degree of relics, already proposed in the local Universe seems to be confirmed but more data are needed.}}
   %{We confirm the relic nature of two out of three galaxies, showing that they formed the majority of their stellar mass at very high redshifts and both their metallicity and [$\alpha$/Fe] are (slightly) super-solar. We thus conclude that KiDS J0224-3143 and J0847+0112 are the first confirmed relics at $0.15<z<0.5$.TO BE CONFIRMED}
  % conclusions heading (optional), leave it empty if necessary 
   %With these results, we validate our strategy for the candidates selection %and for the spectroscopic observations, and confirm our initial estimate that roughly 2/3 of our sample will be most likely confirmed as "bona-fide". This will extend by a factor of $\sim8$ the number of currently confirmed relics at $z<0.5$ 


   \maketitle
%

%%%%%%%%%%%%%%%%% BODY OF PAPER %%%%%%%%%%%%%%%%%%

\section{Introduction}
\label{sec:intro}
A two-phase formation scenario is becoming increasingly favoured to explain the formation of massive early-type galaxies (ETGs; e.g. \citealt{Naab+09,Oser+10,Hilz+13,Rodriguez-Gomez+16, Zibetti20}). 
According to this scenario, a first phase characterised by an intense and fast dissipative series of processes occurring at early cosmic times ($z>2$), forms blue, disk-dominated, and centrally concentrated galaxies \citep{vanderWel+11}
%. Gas accreted early through hierarchical processes settles in the inner disk, and stars are formed 
that form stars 'in-situ' at a high rate (star formation rate, SFR$\ge 10^{3}$ \Msun yr$^{-1}$): %forming 
the so-called blue nuggets \citep{Barro13,Dekel_Burkert14, Zolotov15, Tacchella16}. 
%The turbulent gas continuously feeds the disks which undergo violent instability. The gas is driven into the centre and this dissipative inflow produces massive compact gas-rich bulks that form stars (âin situâ) at a high rate (star formation rate, SFR$\ge 10^{3}$ \Msun yr$^{-1}$),  
Subsequently, the star formation  quenches, for reasons that are still heavily debated (see e.g. \citealt{Martin-Navarro+19} and references therein) and the blue nuggets quickly and passively evolve into compact 'red nuggets' \citep{Daddi+05, Damjanov+09}. 
%The reason for quenching is still under debate. Very recently, \citet{Martin-Navarro+19} have argued that quenching is caused by a combination of central stellar density and halo mass. They interpreted this as a characteristic black hole-to-halo mass ratio, separating quiescent and star forming galaxies. The role of stellar feedback is not expected to be significant given the high stellar masses. 

These high-z red nuggets can typically reach masses similar to those of local giant elliptical galaxies ($\mathrm{M}_{\star}>10^{11}$ \Msun), which indicates that a large portion of the mass is assembled during this first formation phase. However, their sizes are only about a fifth of those of local ETGs of similar masses \citep{Daddi+05, Trujillo+07, vanDokkum08, Werner+18}. 
Thus, during a second phase, at lower redshifts, red nuggets must increase in size, becoming, over billions of years, the present-day giant ETGs. %\footnote{A complication to this scenario arises from the so-called âprogenitor biasâ \citep[e.g.,][]{vanDokkum_Franx01, Carollo+13_quenched}: the later arrival of newly quenched, potentially larger galaxies onto the red sequence, which implies that the average size evolution for the ETG population does not necessarily measure that of the individual galaxies. However, }
%\footnote{We note that \citet{Newman+12} found no significant difference between the rates of size growth at fixed velocity dispersion and mass in a sample of galaxies at $z>1$, suggesting that the role of âprogenitor biasâ \citep[e.g.,][]{vanDokkum_Franx01, Carollo+13_quenched} in interpreting size evolution is not large.}.  

After the second and more time-extended phase, the stars formed during the first phase become the 'core' of today's ETGs and occupy their innermost regions. Stars 'accreted' during the second phase, instead, live preferentially at larger radii (e.g. \citealt{Spavone17}), although it has also been shown that major mergers and accretion of more massive satellites can deposit stars at the centres of the hosts \citep{Amorisco17}. 

Since theoretical evidence suggests that massive (\Mstar$>10^{11}$\Msun) halos at $z = 2$ typically experience no more than one or two major mergers until $z = 0$ \citep{Genel+08, Khochfar_Silk06}, and this is not enough to account for the required mass and size growth; a series of minor dry mergers (i.e. accretions of small gas-poor satellites) seems to be the preferred channel for size evolution, building extended stellar envelopes around compact galaxy cores \citep{Naab+09,vanDokkum+10, Oser+10, Oser+12, Tortora+18_KiDS_DMevol, Stockmann20}. Together with a continuous addition of larger, newly quenched (blue) galaxies to the quiescent population \citep[e.g.][]{vanderwel+09_size_mass, Saglia+10, Carollo+13_quenched, Cassata+13}, this scenario may explain the observed increasing average size of quiescent galaxies with decreasing redshift (e.g. \citealt{vanderWel+14}).  
The early domination of in-situ star formation and the subsequent growth by stellar mergers is also in good agreement with predictions from semi-analytical models \citep[e.g.][]{deLucia+06, Guo+08, Guo+11, Shankar+10} and with observational results in the local Universe \citep{Coccato10,Longobardi15, Iodice17, Pulsoni18, Ferre-Mateu+19, Zibetti20}. %\chiara{ADD MORE REFERENCES}

Unfortunately, in local giant galaxies, the accreted material contaminates the in-situ component\footnote{It has been shown that for more massive galaxies, the accreted component starts to dominate the mass and light budgets at smaller radii than for less massive systems \citep[e.g.][]{Seigar07,Cooper13, Rodriguez-Gomez+16, Spavone17, Spavone20}} that encodes the information about high-z baryonic processes, affecting, along the line-of-sight (LOS), its spatial and orbital distributions, and thus irreversibly limiting our resolving power \citep[e.g.][]{Naab+14, Ferre-Mateu+19}. 
This makes it very difficult to observationally separate star formation history from assembly history. Fortunately, given the stochastic nature of the merging processes, a small fraction
of massive quiescent and very compact galaxies must exist at all redshifts, 
although this fraction should decrease with time.

 %and {\bf heavily depends on the different assumptions and ingredients used to calculate it.

{\sl \emph{Relic galaxies}}, which are red nuggets that survived intact to the present-day Universe without experiencing any merger or interaction, provide a unique opportunity to track the formation of the pristine in-situ galaxy stellar component, which is otherwise mixed with the accreted one in normal massive ETGs \citep{Trujillo+09_superdense}. A detailed study of the stellar population of a large, statistical sample of high-z red nuggets would require prohibitive integration times with the current available facilities. 
Thus, relics are the only systems that allow us to study the physical processes that shaped the mass assembly of massive galaxies in the high-z Universe with the amount of details currently reachable only in the nearby Universe (e.g. high spatial resolution and spectroscopic signal-to-noise ratio). 

However, the exact fraction of such systems at lower redshift critically depends on the physical processes shaping the size and mass evolution of galaxies, such as the relative importance of major and minor galaxy merging (e.g. \citealt{Naab+09,Oser+10, Oser+12, Oogi13,Quilis_Trujillo13,Wellons16}). 
%Moreover, many questions cannot be answered by studying only high-z compact quenched objects: 
Moreover, many questions have still to be answered: how many relics exist in the Universe at each redshift, and how many survive until the present day?  How can they passively evolve through cosmic time without experiencing any interaction? What can we learn on the merging history and the evolution of the most massive ETGs? Is there a physical scenario, predicted by hydrodynamical cosmological simulations, able to explain all the current observational results at all z?  
These  topics are currently heavily debated. 

At high redshift ($z>1$), the number density of compact quiescent galaxies is measured to be in the range %of 
$\rho \sim 10^{-5} $ -- $ 10^{-4}$Mpc$^{-3}$ 
\citep[e.g.][]{Bezanson+13, Saracco+10, Cassata+10,Cassata+13, vanderWel+14}. In the local Universe, so far, only three relics have been spectroscopically confirmed: Mrk1216, PGC032873, and NGC1277, (\citealt{Trujillo+14}, \citealt[][hereafter F17]{Ferre-Mateu+17}). Based on these three confirmed relics, F17 obtained a lower limit to the number density ($\rho$) of relics in the local Universe of $6 \times 10^{-7}{\rm Mpc}^{-3}$. 
This is also consistent with the number density of massive object that formed early and evolved varying their stellar mass by less than %with a stellar mass varying less than
30\% predicted from either %of the
semi-analytical models (Millennium simulations - \citealt{Quilis_Trujillo13}) or hydrodynamical simulations (Illustris - \citealt{Wellons16}). 
%This is also in line with predictions from semi-analytical models based on the Millenium simulation on the number density of massive objects that formed early on in cosmic time and evolved without changing their stellar mass by more than 30\% \citep{Quilis_Trujillo13} and with predictions from the Illustris simulation \citep{Wellons16}.

The three local relics have large rotation velocities ($V\sim 200$ -- $300$ \kms) and high central stellar velocity dispersions ($\sigma_{\star}>300$ \kms);  moreover, their morphology and density profiles match those found in high-z red nuggets and in nearby fossil groups remarkably
well (e.g. NGC~6482, see \citealt{Buote19}). A further confirmation of the genuine relic nature of one of these three systems, NGC1277, came from the uni-modal and uniquely red colour distribution of its globular cluster population \citep{Beasley+18}. For Mrk~1216, \citet{Buote19} found a 
highly concentrated dark matter (DM) halo, reflecting the extremely quiescent and isolated evolution of its stars, and a larger DM fraction within the effective radius.  

From a stellar population point of view, the three objects found by F17 have a single stellar population with super-solar metallicities and old ages out to several effective radii. Their stars are also characterised by a large [Mg/Fe] over-abundance, which is consistent with early and short star formation episodes (within timescales $< 1 \, \rm Gyr$, \citealt{Thomas+05}). 
Finally, the three relics also have a bottom-heavy stellar initial mass function (IMF), with the fraction of low-mass stars being at least a factor of 2 larger than that found in the Milky Way (\citealt{Martin-Navarro+15_IMF_relic}, F17). This result is in good agreement with the proposed tight relation between IMF slope and galaxy velocity dispersion \citep[e.g.][]{Conroy_vanDokkum12a,Cappellari+12, TRN13_SPIDER_IMF, LaBarbera+13_SPIDERVIII_IMF,Spiniello+14}.  
However, while for ETGs with `normal sizes' the bottom-heavy IMF is concentrated only in the very centre  (\citealt{Sarzi+18, Parikh+18, LaBarbera+19}, Barbosa et al. 2020), for the F17 relics the IMF remains virtually invariant with radius. A possible working hypothesis may be that the excess of dwarf stars originates in the first 
%would be then that the excess of dwarf stars could have originated from the first 
phase of the mass assembly, the only one that relics have experienced \citep{Smith2020ARA&A}.

Clearly, in order to validate these various scenarios and test the predictions for the size evolution from different models, more relic statistics are needed, extending the redshift boundaries to bridge the gap with the high-z Universe.
%Clearly, a larger number of relics needs to be confirmed, also extending the redshift boundaries and bridging the gap with the high-z Universe, in order to validate such scenario and test the predictions on galaxy size growth from different galaxy formation models. 
This constitutes one of the main goals of the INvestigating Stellar Population In Relics (\INSPIRE) Project, which we present in this paper. 

The starting point of the \INSPIRE\, project is an approved ESO Large Program (LP; ID: 1104.B-0370, PI: C. Spiniello) of 154 hrs of observations on the the X-Shooter spectrograph (XSH, \citealt{Vernet11}) at the ESO Very Large Telescope. The LP is spread over two  years and targets 52 confirmed ultra-compact massive galaxies (UCMGs) with redshifts $0.1<z<0.5$ and with red colours,  %that are red in colour,
found in the Kilo Degree Survey (KiDS, \citealt{Kuijken11}) Data Release 3 (DR3, \citealt{deJong+17_KiDS_DR3}) from a dedicated project \citep{Tortora+16_compacts_KiDS, Tortora+18_UCMGs, Scognamiglio20}. 

Ultra-compact
massive galaxies are outliers in the mass-size relation \citep[e.g. ][]{Shen+03}
since they have very small sizes (\Reff$\le1.5$ kpc) despite their relatively large stellar masses ($\mathrm{M}_{\star}\ge 8 \times 10^{10}$ \Msun). \footnote{This definition is given in \cite{Tortora+16_compacts_KiDS}. Note however that different studies have adopted slightly different criteria on masses and sizes to define compact galaxies and UCMGs.} They are therefore the perfect relic candidates, to test whether they are populated by a very old single stellar population, formed in-situ, likely with a super-solar [$\alpha$/Fe] and metallicity, and with very little or no sign of subsequent star formation events. Proving this, and thus constraining the stellar population of these objects, is, indeed, the main goal of the \INSPIRE\, Project. 

By means of a detailed stellar population analysis on the sample of 52 KiDS-confirmed UCMGs (relic candidates), once observations are  completed, we will build %\LEt{in the future?}
 the largest sample of relics at $z<0.5$ obtaining a precise relic number density and its evolution in the  low-to-intermediate redshift window. 
The \INSPIRE\, catalogue will represent a fundamental benchmark for ongoing and future cosmological simulations, which will have to be able to reproduce the number density and  characteristics of these systems in order to predict the formation and evolution of local massive galaxies. 
Here in this pilot paper we present the preliminary results on the first three INSPIRE targets, demonstrating that our observation and data analysis strategy works, and therefore that we will be able to assess the true (relic) nature of the candidates. 

\medskip
This paper is organised as follows. In Section~\ref{sec:survey_presentation}, we give an overview of the \INSPIRE\, ESO LP,%\LEt{Please ensure & check throughout the paper that all acronyms and abbreviations are written out at \uline{first mention}, followed by the abbreviation or acronym in parentheses (even if you have already introduced them in the Abstract). After that please use \uline{only the abbreviation}. Instruments, surveys, or facilities do not need an introduction in the Abstract. Please introduce these in the body of the paper unless they are known only by their acronym. See Sect. 5.2.4 of the Language Guide (https://www.aanda.org/for-authors/language-editing/1-introduction) }
 including  details on the sample definition and the current status of the observations and analysis. In Section~\ref{sec:pilot}, we introduce the pilot study %\LEt{program if you're talking about software, otherwise programme   } 
 based on the first three fully observed and analysed galaxies. We describe how we obtained the  final UVB+VIS spectrum for each system and how we calculated the line-of-sight velocity distribution and the stellar population parameters from full spectral fitting. Finally, in Section~\ref{sec:results} we present the results of this study, including a comparison to F17, and in Section~\ref{sec:conclusions} we draw our conclusions. Throughout the paper, we assume a standard $\Lambda$CDM cosmology with $H_0=69.6$ \kms Mpc$^{-1}$, $\Omega_{\mathrm{vac}} = 0.714$ and $\Omega_{\mathrm{M}} = 0.286$ \citep{Bennett14}.

\section{The ESO Large Program INSPIRE}
\label{sec:survey_presentation}
The \INSPIRE\, project aims to characterise the kinematical, dynamical, and stellar population properties of 52 confirmed UCMGs at redshift $0.1<z<0.5$ with unprecedented detail. These objects have been selected because their multi-band photometry is consistent with having old stellar populations, from multi-band photometry (see Sect.~\ref{subsec:sample} for more details). 

\INSPIRE\ has been designed to obtain spectra with a signal-to-noise (S/N)  high enough  
%and it is suitable 
for detailed spectroscopic stellar population analysis, which is the only way to confirm the relic nature of the stellar populations of these UCMGs. Furthermore, the large wavelength range of XSH is particularly suitable to infer the IMF slope at its low-mass end, which represents one of the main pillars of the project. In fact, a very wide wavelength interval (also covering the near-infrared where cold dwarf stars contribute up to 30\% of the stellar light \citep{Worthey+94}) is crucial in breaking the stellar population degeneracies between variations in age, metallicity, elemental abundances, and IMF slope \citep{Spiniello+14, LaBarbera+19}. % \footnote{The IMF estimates will be presented in a forthcoming paper of this series.}. %Furthermore, with the collecting power of the VLT, we can reach high SNR, suitable for detailed spectroscopic stellar population analysis, crucial to confirm the relic nature of the stellar populations of these UCMGs.
%Thanks to high signal-to-noise (SNR) XSH spectra, with the \INSPIRE\, project, we will  characterise the kinematical, dynamical and stellar population properties of 52 confirmed UCMGs at redshift $0.1<z<0.5$ in great details. These objects have been selected because consistent with having old stellar populations, from multi-band photometry (see Sec.~\ref{subsec:sample} for more details).
%The need for XSH is motivated by the fact that one of the main goals of the \INSPIRE\, is to infer the IMF slope at its low mass end. This is only possible if a very wide wavelength range is available, covering also the near infrared where cold dwarf stars contribute by up to 30\% of the stellar light \citep{Worthey+94}\footnote{The IMF estimates will be presented in a forthcoming paper of this series.}.  Furthermore, with the collecting power of the VLT, we can reach high SNR, suitable for detailed spectroscopic stellar population analysis, crucial to confirm the relic nature of the stellar populations of these UCMGs.  

In the following sections, we describe how the relic candidates targeted by the \INSPIRE\, project have been selected, and we give some details regarding the strategy and the current status of the observations. 
The full sample will be presented in the first \INSPIRE\, Data Release, expected at the beginning of 2021,  where we will also describe in more detail the main goals and scientific objectives of the project.  Constraints on the IMF will be presented in a dedicated forthcoming publication,  since this requires very careful treatment of the telluric atmospheric absorption lines in the infrared, and possibly a more sophisticated approach to the stellar population analysis.

\subsection{Sample definition}
\label{subsec:sample}
%These UCMGs are part of a large catalog that
With a dedicated collaboration project, our team has started a systematic census of UCMGs in the redshift range $0<z<0.5$ within the KiDS footprint. 
Thanks to its exquisite image quality (angular scale of $0.21\arcsec$/pixel and a median r-band seeing of $\sim0.65\arcsec$; \citealt{Kuijken11,deJong+15_KiDS_paperI}), KiDS is optimally suited for the purpose of carrying on a complete census of these ultra-compact systems.   %up to $z\sim0.5$,  
%Thanks to a dedicated project, as part of the Kilo Degree Survey \citep{Kuijken11} collaboration \citep[][hereafter T16]{Tortora+16_compacts_KiDS}, our team  
%
%We assembled, in the last few years, t and a multi-site, multi-telescope spectroscopic follow-up campaign (T18 and S20), which we describe in more details in the next section.  
%In the last few years, our team
In \citet[][hereafter T18]{Tortora+18_UCMGs}, we presented $\sim1000$ candidates, with \Reff$\le1.5$ kpc and $\mathrm{M}_{\star}\ge 8 \times 10^{10}$, collected from the first $333$ $\mathrm{deg}^{2}$ of KiDS (DR3, \citealt{deJong+17_KiDS_DR3}).

\begin{figure}
    \centering
    \includegraphics[width=\columnwidth]{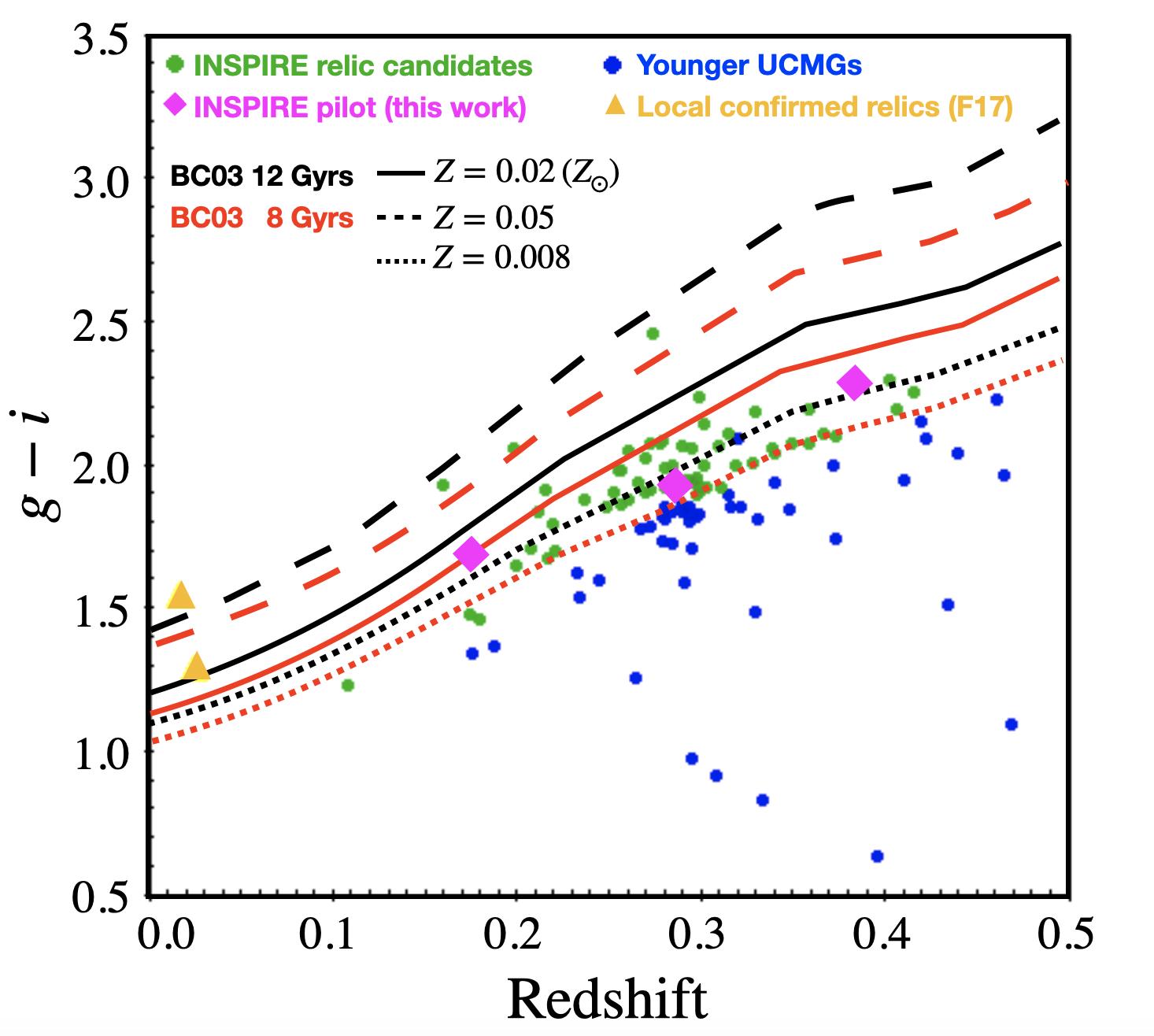}
    \caption{$g-i$ colour plotted versus redshift for the  spectroscopically confirmed UCMGs in KiDS ($\sim100$, T18, S20). The relic candidates, targets of the XSH \INSPIRE\, LP, are plotted as green points. The three objects, which are the focus of this study, are plotted as magenta diamonds. Blue points are UCMGs from T18 and S20 consistent with younger stellar ages. Two out of the three confirmed relics of F17 are plotted as yellow triangles (colours are taken from the SDSS DR16). Lines show single-burst old stellar population models \citep{ Bruzual+03} with solar (Z$=0.02$, solid lines),  sub-solar (Z$=0.008$, dotted), and super-solar (Z$=0.05$, dashed) metallicity and ages of 12 Gyr (black) and 8 Gyr (red).} 
    \label{fig:old_ages_color}
\end{figure}

We also recently started a programme to spectroscopically confirm these candidates using public %low SNR
spectra from the literature (2DFLenS, SDSS, and GAMA) as well as %\LEt{not sure what you're trying to say from here. 'and' is repeated a lot.}
new spectroscopy. We currently confirmed $\sim100$ of them (T18 and  \citealt[][ hereafter S20]{Scognamiglio20}). Although these spectra allowed us to estimate the redshifts and thus confirm the extragalactic, ultra-compact, and massive nature of these objects, their S/Ns were too low to use them for any stellar population analysis.

%we have also already spectroscopically confirmed $\sim100$ of them, getting their redshifts through public low SNR spectra from 2DFLenS, SDSS and GAMA and new spectroscopy \citep[][ hereafter T18 and S20]{Tortora+18_UCMGs, Scognamiglio20}.  

We built the largest sample of spectroscopically confirmed UCMGs at redshifts $0.1<z<0.5$. This redshift range is an important link between the very small sample of ultra-compact galaxies in the local Universe and the large population of high-redshift red nuggets. In particular, it represents a sweet spot where it is possible to survey larger volumes with large-sky photometric surveys but also to target the candidates with spectroscopic surveys/facilities \citep{Damjanov+13_compacts, Damjanov+14_compacts, Damjanov+15_compacts, Buitrago+18_compacts}, as they are brighter than the higher-z counterparts.  Currently, this is the only way to enable statistical studies of structural, dynamical, and environmental properties of UCMGs \citep[see e.g.][]{Damjanov+15_env_compacts, Tortora20}. 

Among the $\sim100$ spectroscopically confirmed UCMGs, half of them have broad-band colours consistent with old stellar populations, typical of red and dead galaxies.
%show convincing evidence, from broad-band colours,  %and low resolution spectroscopy,  of old stellar populations,  typical of red and dead galaxies. %\LEt{again, not sure what you mean. They show evidence that comes from broadband colours? The evidence is for old stellar populations, or the broadband colours? the populations are typical of dead galaxies? or the colours? or the evidence...??} 
 To show this, and thus select the optimal relic candidates, we compared the UCMG's $g-i$ colours to those of single-burst stellar population models from \citet{Bruzual+03}. We then only selected objects %We then retrieve everything 
 consistent with a formation epoch that occurred at least 8 Gyr ago (also allowing for super- and sub-solar metallicities), 
 %Gyrs ago (allowing also for sub-solar metallicity, Z$=0.4$Z$_{\odot}$, which would make the color bluer), 
 as visible from Figure~\ref{fig:old_ages_color}, where the $g-i$ colour (in the observed-frame) is plotted against the redshift. Blue points in the figure are the $\sim 100$ UCMGs for which we have spectroscopic redshifts available. The systems selected for the \INSPIRE\, ESO LP are plotted in green, while the three magenta diamonds show the position of the three systems that we study in this pilot paper, as described in Sect.~\ref{sec:pilot}. Finally, the yellow triangles show the positions of two of the three confirmed local relics in F17 (NGC1277 and PGC032873) that are also in the SDSS footprint. For these objects, the $g-i$ colour is computed using the 'modelMag' from the SDSS DR16 \citep{Ahumada20_sdssdr16}. Unfortunately, the third relic in F17 (Mrk 1216) is not in the SDSS footprint and thus we do not have magnitudes available for this object. 

The 52 \INSPIRE\, targets span a very wide range in right ascension and declination (i.e. the optimum observability from Paranal spread over a full year) and are all brighter than 20 mag in the $r$-band. 
Since the effective radii of the galaxies are very small and there is no possibility to spatially resolve them from the ground (without adaptive optics), we required relatively 'poor' conditions for the observations (seeing up to $1.2\arcsec$, CLR nights, grey lunar phase up to a Moon FLI of 0.6). This increased the efficiency of the service mode under a LP. 

For all 52 systems, we have multi-band optical photometry available from the KiDS (DR3) catalogue, structural parameters derived in \citet{Roy+18} and spectroscopic redshifts retrieved from the literature (from SDSS, \citealt{Ahn+12_SDSS_DR9, Ahn+14_SDSS_DR10}; 2dFLenS, \citealt{Blake+16_2dflens};    
GAMA, \citealt{Driver+11_GAMA}) or calculated through new spectroscopic observations presented in previous papers of our group (T18 and S20).  
%photometric redshifts obtained with the Multi Layer Perceptron with Quasi Newton Algorithm (MLPQNA, \citealt{Brescia+13, Cavuoti+17_METAPHOR}) method and presented in \citet{Cavuoti+15_KIDS_I, Cavuoti+17_KiDS}  as well as structural parameters \citep{Roy+18}. 

These UCMGs are the optimal relic candidates: they have been confirmed to be massive (M$_{\star}\ge 6 \times 10^{10}$ \Msun) and compact (\Reff $\,\le 2$ kpc) extragalactic objects and are consistent with having passive stellar populations. 
We note that the size and mass selection criteria used here to select relic candidates are slightly more relaxed than those used for UCMGs in T18 and S20. In fact, these systems were originally selected to have \Reff $\,\le 1.5$ kpc and M$_{\star}\ge 8 \times 10^{10}$ \Msun on the basis of their photometric redshifts. However, after recomputing these quantities using the spectroscopic redshifts instead, some sizes (stellar masses) resulted to be slightly larger (smaller) for some objects, which could still be interesting, on the basis of their stellar population properties. Because the validation of UCMGs is based merely on the measured values, without taking into account any uncertainties (which in some extreme cases, i.e. for very small radii, can be as high as 50-80\%: see e.g. T18), and we need to build a statistically large sample, we relaxed both the compactness and massiveness thresholds.  Thus we selected, for the XSH Proposal, galaxies with \Reff $<2$ kpc\footnote{We note that this also better aligns our size criterion to that used for the three confirmed relics of F17 and other papers in the literature.} and M$_{\star}\ge 6 \times 10^{10}$ \Msun. 

\begin{figure}
    \centering
    %\vspace{0.1cm}
    \includegraphics[width=\columnwidth]{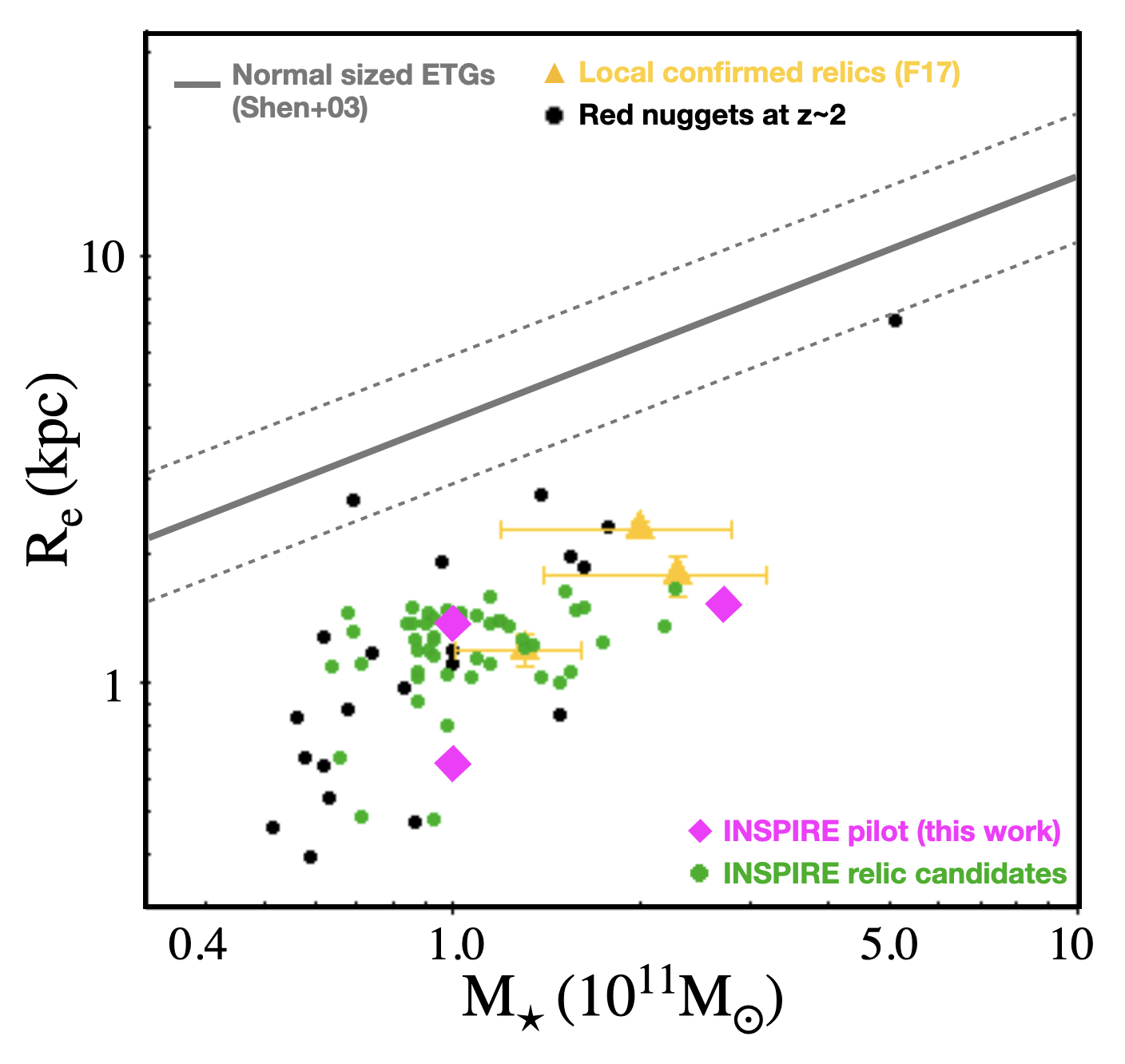}
    \caption{Stellar mass-size relation for high-z nuggets (black points, from \citealt{Szomoru+12}), the F17 local relics (yellow triangles), and the relic candidates targeted by \INSPIRE (green points and magenta diamonds), compared to the relation for normal SDSS ETGs (grey line, from \citealt{Shen+03}). By selection, all UCMGs are clear outliers of the local stellar mass-size relation, and high- and low-z systems lie on the same region of the diagram. }
    \label{fig:mass_size}
\end{figure}

All these objects  %circularized effective radii well constrained, thanks to exquisite image quality from KiDS. They also 
have similar or slightly smaller sizes and masses than the three confirmed local relics of F17\footnote{We note that by adopting an upper size threshold of \Reff = 1.5 kpc, T18 selects only the most compact objects. By relaxing this criterion, we would most likely have found a few more objects as massive as the three F17 galaxies.} and high-z red nuggets, and they lie below the mass-size relation defined from normal-size ETGs of equal masses at similar redshift \citep{Shen+03, Roy+18}, as can be seen from Figure~\ref{fig:mass_size}.  
%Also, the stellar mass densities of the three local relics do not resemble any of the average profiles found in local Universe galaxies, being significantly denser in the inner regions compared to galaxies of similar mass but with more normal sizes, while they match almost perfectly those of the high redshift massive population 
%\citep[e.g.][] F17, 81) 
%Morphologically speaking, 
In terms of their morphology, like the confirmed local relics, the majority of the \INSPIRE\, targets show compact elongated disk-like shapes typical of the massive galaxy population at redshifts $z>2$ (F17, \citealt{vanderWel+11}). 
Finally, the KiDS images, which usually go as deep as 26 mag/arcsec$^2$ in the r-band surface brightness (see Fig.~3 in T18), do not seem to reveal signatures of tidal tails, asymmetries, or any other features that could indicate any past or current interaction or stripping (see Fig.~2 in T18 and Fig.~1 in S20).\footnote{One of our targets, J0847+0112, is in the wide layer of the Hyper Suprime-Cam Subaru Strategic Program of the GAMA survey (\url{https://hsc.mtk.nao.ac.jp/ssp/survey/}). Also in this case, from the publicly available deep multi-band images ($r\sim26$ mag) of their DR2, we do not detect any prominent structure. However, a hint of a very faint small tidal feature east of the galaxy can be seen in the i-band. Further investigation on the possible nature of this feature (likely some background sources at higher redshift) will be performed in the future.}

\subsection{Observation strategy and current status}
\label{subsec:data}
The observing strategy was set to be the same for all the systems: 
we chose slit widths of 1.6$\arcsec$, 1.5$\arcsec,$ and 1.5$\arcsec$, respectively
for UVB, VIS, and NIR to ensure minimal slit loss. The XSH slit length is 11$\arcsec$, 
which ensures that we cover within the slit %within the slit of 
a region far enough from the galaxy, allowing us to perform a proper sky subtraction. For this purpose, we used a dithering scheme (in the NODDING MODE) with multiple frames where the galaxy 
is offset by a small amount from the centre of the slit (jitter box of 0.2$\arcsec$ along the slit width and a shift of 2$\arcsec$ along the slit length). The position angle (PA) was always aligned with the major 
axis of the galaxies, which was inferred from the $r$-band KiDS images.

We scheduled from 1 to 4 OBs (of one hour each) per system, based on their surface brightness luminosities (obtained performing a Se\'rsic fit) averaged within the effective radius ($\langle\mu_{\mathrm {e}}\rangle$) and aperture magnitudes in the $r$-band (obtained from the KiDS catalogue). 
To determine the integration times for each target, we used the ESO XSH exposure time calculator with a template spectrum of an elliptical galaxy and testing both an extended spatial distribution and a point-like source, since our objects are very compact on the sky (r-band \Reff $\,\sim 0.2-0.6\arcsec$). 
In particular, the integration times are driven by the quality of the spectra we need to reach in terms of S/N, in order to %very
%accurately 
achieve the scientific goal of accurately constraining the integrated stellar population parameters, including a precise estimate of the stellar IMF slope. 
Based on our previous experience and on literature results \citep{CidFernandes14, Ge+18, Costantin19}, an integrated S/N $\ge15$ per \AA \, allows the recovery of stellar age, metallicity, and [$\alpha$/Fe] with uncertainties smaller than 0.1 dex (using both full-spectral fitting with single stellar population modelling and line index measurements), and thus the confirmation of the relic's nature. However, a much higher S/N ($>40$ per \AA, based on our experience) is required, at least in the NIR, to also properly infer the low-mass end slope of the IMF (see e.g. \citealt{LaBarbera+19}).   
For this purpose, XSH is the ideal instrument since its very large wavelength baseline allows one to 
%In particular, for this purpose and as we already noted, XSH is the ideal instrument for achieving our aims. This is because a very large wavelength baseline
%is required to
break the degeneracies between the stellar population parameters and at the same time obtain a secure and precise constraint (with 
10\% accuracy) on the low-mass end slope of the IMF. 
%The IMF has been inferred to be dwarf-rich (or bottom-heavy) in the three local relics of F17, up to few effective radii (\Reff). In contrast, in very massive local ETGs, the dwarf-rich IMF is restricted to the innermost region (see e.g. \citealt{Sarzi+18}). 
We will constrain the IMF, fitting the full UVB-to-NIR wavelength range (at least for the objects with the highest S/N spectra) in a forthcoming paper of this series since this %is because it
requires a more careful analysis of the infrared spectrum. %, which is not analysed here. 
In this way, we will be able to draw a scenario to explain the different observations derived for relics and normal-size galaxies. At the time of writing, the first period of observations (ESO-P104) has concluded, and 42 hours of observations have been carried out on 17 galaxies. 

The first \INSPIRE\, Data Release, and the associated dedicated paper, comprising all the spectroscopic observations carried on during the first semester (running from October 2019 until March 2020) is planned for the first half of 2021. %, covering all galaxies with already completed observations, once all the data will be reduced and analyzed.   
Meanwhile, to test our observations, data reduction, and analysis strategy, %and, at the same time introducing the on-going survey, 
we present here the initial results for a pilot study focusing on three systems, representative of the whole sample, and for which the spectroscopic observations have already been completed: KiDS J0224-3143, KiDS J0314-3215, and KiDS J0847+0112. 

Another necessary step, which is the subject of ongoing work, is the correction of the telluric atmospheric absorption lines, which severely affect the red part of the VIS and the NIR arms.  
Therefore, for the moment, we limit the analysis to the UVB and blue part of the VIS (from $\sim350$ to $\sim650$ nm), which is sufficient to derive the line-of-sight velocity distribution (LOSVD)\footnote{Here we limit the analysis  %\LEt{limit what?}
to the first two moments of the LOSVD, namely velocity and velocity dispersion, but we note that the final S/N of the spectra, as well as the covered wavelength range are good enough to get a reliable estimate of the higher Gauss-Hermite moments, $h3$ and $h4$, representing skewness and kurtosis, respectively. These will be presented in the data release 1 paper, already in preparation.} and to break the stellar age-metallicity degeneracy, and thus constrain the stellar age and metallicity from full spectral fitting and $\alpha$ abundance from line indices with uncertainties smaller than 10\%.

\begin{table*}   
\caption{First three columns show the ID, right ascension, and declination of the three relic candidates that we study in detail in this paper as part of the pilot study of the \INSPIRE\, project. 
The central block lists  morpho-photometric characteristics of the objects derived from the KiDS $r$-band images shown in Figure~\ref{fig:cutouts}. The last three columns list quantities derived in S20.}
%ID, right ascension, declination, position angle of the slit, final exposure time on targets and characteristics of the three relic candidates that we study in details in this paper, as pilot program of the whole \INSPIRE\, project. See text for more details on how these quantities have been derived.}

\label{tab:sample_previous}
\centering
\begin{tabular}{lrr|ccccc|ccc}
\hline
\hline
  \multicolumn{1}{c}{KiDS} &
  \multicolumn{1}{c}{RA$_{\mathrm{J}2000}$} &
  \multicolumn{1}{c|}{DEC$_{\mathrm{J}2000}$} &
  \multicolumn{1}{c}{$n_{r}$} &
  \multicolumn{1}{c}{$q_{r}$} &
  \multicolumn{1}{c}{mag$_r$} &
  \multicolumn{1}{c}{\mue$_{r}$} &
\multicolumn{1}{c|}{\Reff$_{r}$} &
%  \multicolumn{1}{c}{$\langle\mathrm{R}_{\mathrm{e}}\rangle\,$} &
  \multicolumn{1}{c}{$z_{\mathrm{spec}}$} &
  \multicolumn{1}{c}{$\langle\mathrm{R}_{\mathrm{e}}\rangle\,$} &
  \multicolumn{1}{c}{M$_{\star}$} \\
  \multicolumn{1}{c}{ } &
  \multicolumn{1}{c}{(deg)} &
  \multicolumn{1}{c|}{(deg)} &
  \multicolumn{1}{c}{ } &
  \multicolumn{1}{c}{ } &
  \multicolumn{1}{c}{(AB)} &
  \multicolumn{1}{c}{(AB)} &
  \multicolumn{1}{c|}{($\arcsec$)} &
% \multicolumn{1}{c}{($\arcsec$)} &
  \multicolumn{1}{c}{ } &
  \multicolumn{1}{c}{(kpc)} &
  \multicolumn{1}{c}{($10^{11}$ M$_{\odot}$)} \\
\hline
J0224-3143 & 36.090266 & -31.724492 &  6.5  & 0.39 & 19.25 & 17.98 & 0.25 & 0.3853 & 1.54 &  2.7 \\
  %11.43 \\
J0314-3215 & 48.594256 & -32.263268 &  6.36 & 0.38 & 19.57 & 16.98 & 0.15 & 0.2888 &  0.65  & 1.0 \\ %11.00 \\
J0847+0112 & 131.911239 & +1.205713 & 4.38 & 0.25 & 18.41 & 18.69 &  0.47 & 0.17636$^*$ & 1.36  & 0.98 \\ %10.99 \\
\hline
\hline
\end{tabular}
\begin{flushright}
\footnotesize{$^*$ Derived from GAMA.}
\end{flushright}
\end{table*}

\begin{table}   
\caption{Summary of spectroscopic observations. We list: ID, position angle of the slit, final exposure time on targets, median seeing, and inferred redshift of the three relic candidates studied in this paper.}
\label{tab:sample_xsh}
\centering
\begin{tabular}{lcccc}
\hline
\hline
  \multicolumn{1}{c}{KiDS} &
%  \multicolumn{1}{c}{RA$_{\mathrm{J}2000}$} &
%  \multicolumn{1}{c}{DEC$_{\mathrm{J}2000}$} &
  \multicolumn{1}{c}{P.A.} &
  \multicolumn{1}{c}{Exp.T.} &
  \multicolumn{1}{c}{Median} &
\multicolumn{1}{c}{$z_{\mathrm{XSH}}$} \\
  \multicolumn{1}{c}{ } &
%  \multicolumn{1}{c}{(deg)} &
%  \multicolumn{1}{c}{(deg)} &
  \multicolumn{1}{c}{(degrees)} &
  \multicolumn{1}{c}{(sec.)} &
  \multicolumn{1}{c}{Seeing ($\arcsec$)} \\
  \multicolumn{1}{c}{ } \\
\hline
J0224-3143 & %36.090266 & -31.724492 &
310.9 & 11240 & 1.2 & 0.3841 \\
J0314-3215 & %48.594256 & -32.263268 &
263.4 & 8430 & 0.8 & 0.2874 \\
J0847+0112 & %131.911239 & +1.205713 &
314.4 & 5620 & 1.1 & 0.1764 \\
\hline
\hline
\end{tabular}
\end{table}

\medskip
\medskip
\section{The INSPIRE pilot study: KiDS J0224-3143, KiDS J0314-3215, and KiDS J0847+0112}
\label{sec:pilot}
Observations for KiDS J0224-3143, KiDS J0314-3215, and KiDS J0847+0112, to which we gave priority to test our strategy, data reduction, and analysis routines, were completed in 2019. %These three objects have been used to test the observational strategy, as well as the data reduction and analysis routines. 
We highlight that these three objects are representative of the whole sample as they span a relatively wide range in redshift, stellar masses, sizes and magnitudes. In particular, KiDS J0224-3143 is the most massive system in our sample (M$_{\star} = 2.7\times10^{11}$\Msun), and among the galaxies with the highest redshift ($z=0.3841\pm0.0001$, corresponding to a Universe %\LEt{only capitalise for \textit{our} Universe, please check throughout}
that is 9.5 Gyr old.\footnote{The age of the Universe has been calculated using Ned Wright's Javascript Cosmology Calculator (\url{http://www.astro.ucla.edu/\%7Ewright/CosmoCalc.html}) %{Ned Wright's Javascript Cosmology Calculator}
and assuming general cosmology with $\Omega_{\mathrm{vac}} = 0.714.$})  KiDS J0314-3215 is slightly less massive and closer ($z=0.2874\pm0.0001$, when the Universe was 10.4 Gyr old), but it is one of the smallest systems  in the current sample, with an effective radius of only \Reff$=0.65$ kpc, corresponding roughly to 0.16" on the sky (a single pixel in XHS). Finally, KiDS J0847+0112 is brighter than the other two (mag$_r = 18.41$ mag), % and \mue$=18.69$ mag/arcsec$^{2}$) 
and it is one of the \INSPIRE\, galaxies with the lowest redshifts ($z=0.1764\pm0.0001$, when the Universe was 11.5 Gyr old). 

\begin{figure*}
\includegraphics[width=18.5cm]{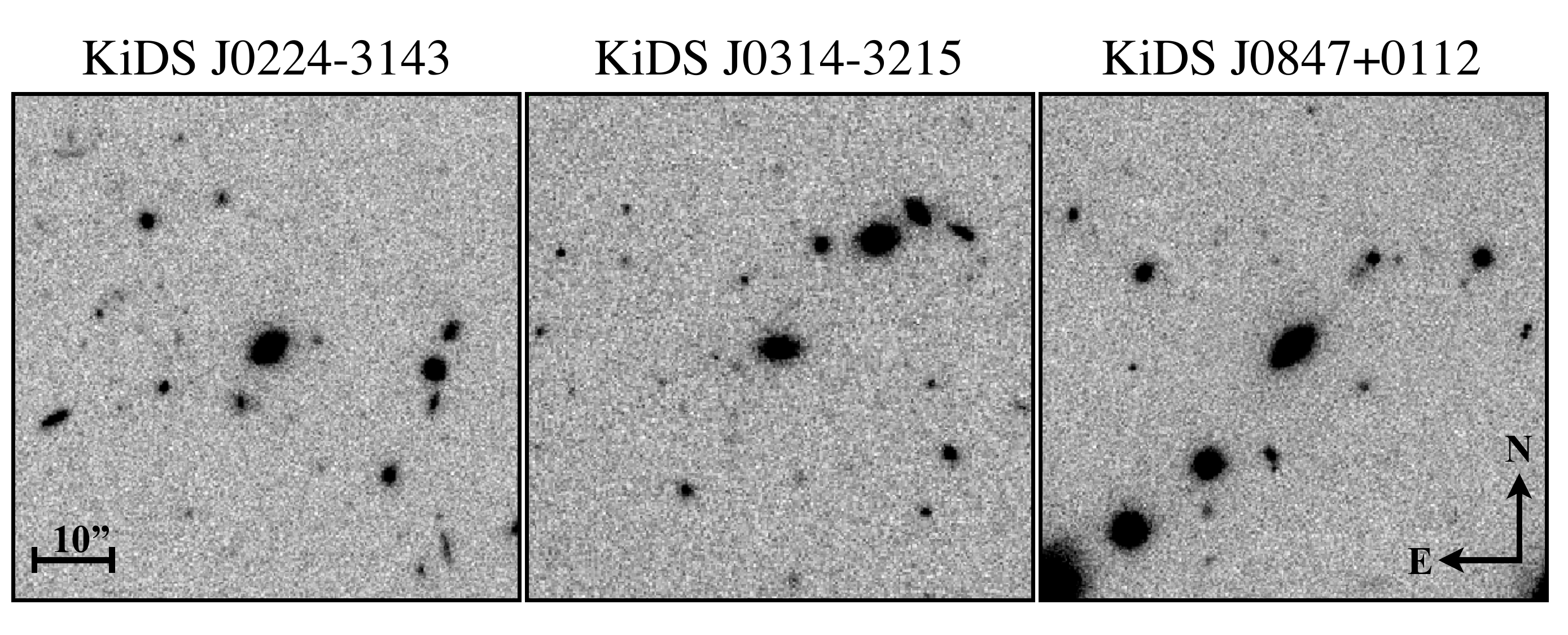}
\caption{Cutouts of $1\arcmin\times1\arcmin$ of the KiDS $r$-band images for the three relic candidates that we study in this paper. Spatial scale and image orientation are common to all the cutouts and are shown in the first and third panels, respectively.}
\label{fig:cutouts}
\end{figure*}

The main morpho-photometric characteristics of these three objects are listed in  Table~\ref{tab:sample_previous}.  %and Table~\ref{tab:sample_xsh}. 
The first block (separated by vertical lines) gives the KiDS ID and the coordinates of the objects; the second block lists a number of properties derived from the $r$-band KiDS images (shown in Figure~\ref{fig:cutouts}), which is the band with the best resolution and seeing from KiDS. 
In particular, we list the total magnitude (MAG\_AUTO, in the KiDS catalogue, which is obtained from \textsc{sextractor},  \citealt{Bertin_Arnouts96_SEx}) and the average surface brightness within the effective radius in the $r-$band. We also give the effective radius, \Reff, S\'ersic index, $n$, and axis ratio, $q$, always in the same band and calculated fitting a PSF convolved S\'ersic profile to the images with the code \textsc{2dphot} \citep{LaBarbera_08_2DPHOT}. 
These latter two quantities demonstrate that the objects are bulge-dominated but elongated, thus hinting at a disk-like shape (see also T16, T18, S20). 
Finally, in the last block, we provide the the spectroscopic redshifts, the median (of the values for $g$, $r$ and $i$ bands) effective radii converted into kpc, and the associated stellar masses as derived in S20.   %The total stellar mass are calculated with the code \textsc{le phare} \citep{Arnouts+99, Ilbert+06}.}

Quantities that refer to the XSH observations are given in Table~\ref{tab:sample_xsh}.  These include the position angle, PA, which is always aligned with the photometric major axis, the final exposure time on target,  %(we obtained a different number of OBs for the different objects, ranging from 2 to 4), 
the median seeing of the observations, and the spectroscopic redshift directly inferred from the XSH final spectra.  
The redshifts we derive from these new spectra are qualitatively consistent (within 5\%) with those derived in T18 (and reported again in S20) for the first two galaxies, but more reliable thanks to the better quality of the new spectra. For J0847+0112, the redshift estimate we compute is  consistent with the one inferred by the GAMA Survey ($z=0.17636$). 

%In Table~\ref{tab:sample_previous} we give instead some 

\subsection{Data reduction and extraction of the 1D spectra}
\label{subsec:1d-extr}
The data has been reduced with the ESO XSH pipeline (v3.3.0) under the EsoReflex workflow \citep{Freudling+13} until the production of the 2D flux-calibrated spectra. 
The extraction of the 1D ones, which is described in more detail below, was instead performed making use of IDL and Python routines specifically developed for the purpose. 

As mentioned before, the spatial resolution of XSH is not good enough to resolve such compact objects.  In fact, in all the cases, the effective radii in arcseconds ($0.25\arcsec$, $0.15\arcsec,$ and $0.47\arcsec$ in the $r$-band  for KIDS J0224-3143, KiDS J0314-3215, and KiDS J0847+0112, respectively) are much smaller than the seeing (which varied from $\sim0.8\arcsec$ to $\sim1.2\arcsec$ for these three systems). In the extreme case of KiDS J0314-3215, the half-light radius is comparable to a %even smaller than a
single detector pixel. 
This means that the spectra are completely seeing dominated.
Thus, the most meaningful approach to compare the galaxies with each other is to try to extract the spectra from an aperture that contains more or less the same fraction of light for the three objects, given the seeing during observations and their surface brightness profiles.  
We therefore used the following procedure: i) we extracted the surface brightness profiles of the three objects directly from the 2D spectra (collapsing the 2D frame in the dispersion direction and assuming spherical symmetry); ii) we integrated these profiles up to the radius where the flux reaches the null value, which we consider as the radius containing the total light; iii) we computed the same integral for different apertures (i.e. integrating only up %\LEt{??}
to smaller and smaller radii); and iv) we took the ratio between these integrals and the total one to obtain the fraction of light as a function of the aperture radius.  
%One could think that this means that we are covering a very different physical radius for the three objects, since in terms of the effective radii we span from  
%{\bf This means that we cover roughly 2.5 \Reff for J0314-3215, to 1.6 \Reff for J0224-3143 and 0.85 \Reff for J0847+0112. However, 

We find that extracting from a fixed aperture of $r=0.4\arcsec$, corresponding to the %meaning from the  
five innermost central pixels, roughly corresponds to 30\% of the total light for all three objects.\footnote{This percentage varies from 0.27 to 0.35 according to the different seeing values of the various OBs.} With this aperture, we maximise the S/Ns of the extracted spectra and make sure to probe a region dominated by baryons. 
This strategy will be revised in the forthcoming papers of this series, where we will also test an optimal extraction procedure, which should allow us to further increase the final S/N of the 1D spectrum \citep{Naylor+98} by $\sim10$\%. 

%study the fraction of light as a function of the aperture and most likely extracting the kinematics and the stellar populations of the galaxies for different choices, but always trying to have a sufficiently high SNR.

%(e.g. by using the optimal extraction approach described in \citet{Horne+86} to obtain spectra all covering a similar spatial aperture). 

Before extracting the spectra, we applied a simple clipping procedure to identify and interpolate the bad pixels, cosmic ray, and sky residuals directly on the 2D frames of each single exposure for each arm. On average, we only interpolate $\sim4$\% of the total pixels.   
We then performed the 1D extraction for each OB, in each arm separately, over an aperture of $\pm0.40\arcsec$ centred on the photometric centre of the object (the pixel with the largest flux), which was chosen to maximise the S/N of the final spectra. 
We also ran the code \mbox{{\tt MOLECFIT}} \citep{Smette15, Kausch15} on the VIS arm.\footnote{We note that the telluric correction we applied here must be further improved, especially at wavelengths redder than 7000\AA, which are not used in this paper. A more accurate correction will be performed in the redder part of the VIS and in the NIR in forthcoming publications presenting the full \INSPIRE\, data and the inference on the IMF.}   
Then, we obtained a final spectrum for each system in each arm, summing the spectra relative to different OBs, weighted for their corresponding errors. Finally, after extracting and summing, we joined together the UVB and VIS final spectra for each galaxy, as described in the next section. 

%The resulting 1D spectra are plotted in Figure~\ref{fig:final_1dspec}, at restframe. Some of the main stellar absorption features that we will use in the following sections to calculate the LOSVD and infer the stellar population parameters are highlighted with red vertical dashed lines. The region where the UVB and VIS arms were joined is shown with a grey shaded box. To avoid possible problems of flux calibration at the boundaries of the two arms, we manually assign to these pixels a very large error, so that they will not be considered in the stellar template fitting procedure described in the next sections. 
%Similarly,  we mask from the fit regions affected by telluric absorption lines (purple vertical areas) and sky-subtraction residuals, as we will describe in more details in Section~\ref{sec:kinematics}. 

\subsection{UVB and VIS combination}
\label{subsec:uvb+vis}
\begin{figure}
    \centering
    \includegraphics[width=9cm]{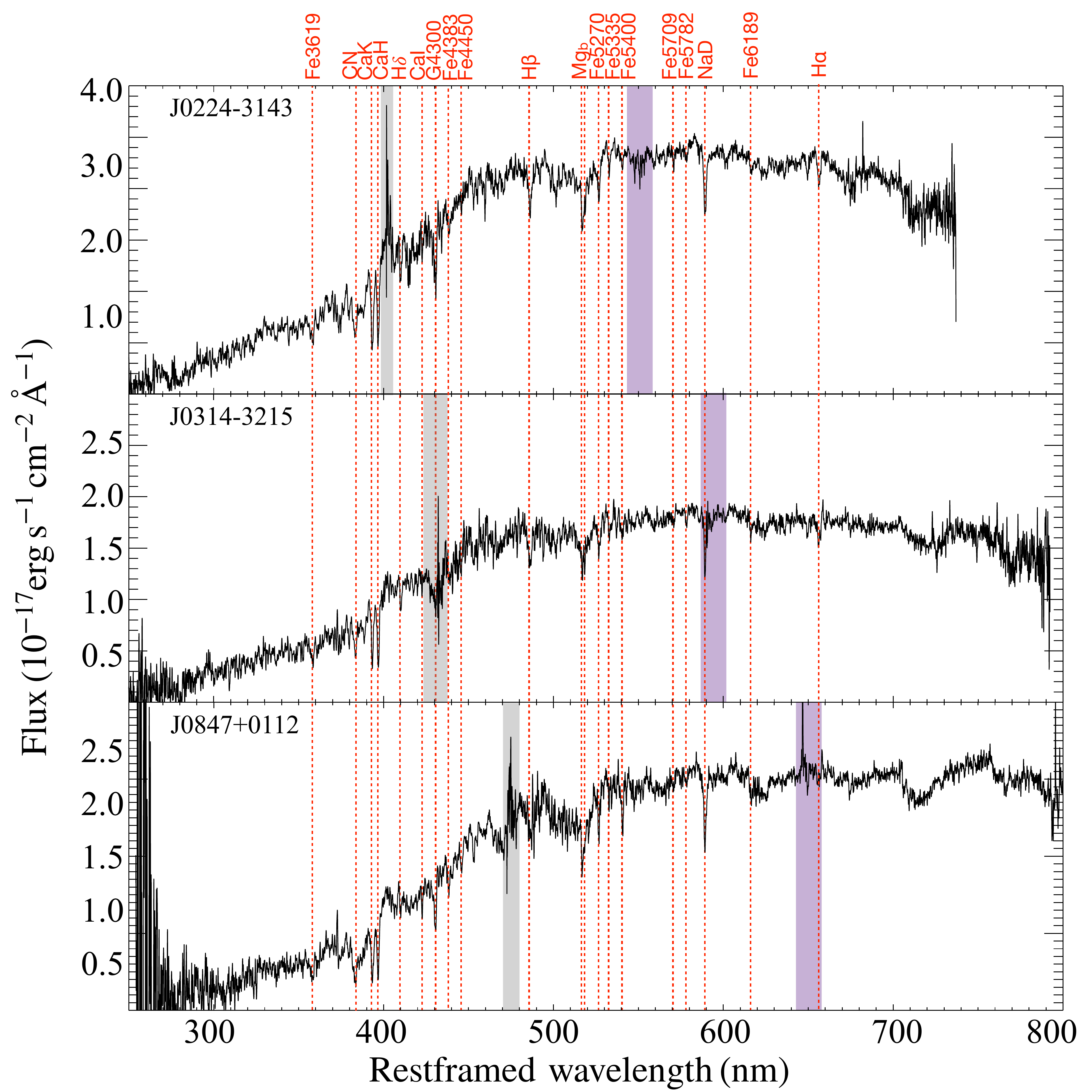}
    \caption{Final 1D spectra extracted from an aperture of five pixels ($\pm 0.4\arcsec$ from the photometric centre) for the three systems presented in this pilot paper, from the one at highest redshift (KiDS J0224-3143, top panel) to the one at the lowest redshift (KiDS J0847+0112, bottom panel). The spectra are %are smoothed to a resolution of 1\AA\,  
    plotted in rest-framed wavelength and fluxes (in units of $10^{-17}$ erg s$^{-1}$cm$^{-2}$\AA$^{-1}$). Red dashed lines highlight the stellar absorption lines, typical of ETGs, that are used to derive stellar kinematics and population parameters. Grey vertical shaded boxes show the regions where the UVB and VIS spectra were joined for each system. The purple boxes show regions contaminated by telluric lines that we corrected with \mbox{{\tt MOLECFIT}}. }
    \label{fig:final_1dspec}
\end{figure}

To correctly infer the stellar population parameters, it is crucial to analyse a sufficiently large wavelength range. This is the only way to break the age-metallicity \citep{Worthey+94}
and metallicity-[$\alpha$/Fe] degeneracy (e.g. \citealt{Spiniello+14}) and study how these stellar population parameters  correlate with galaxy masses \citep{Spinrad62, Frogel78, Faber80, Cenarro+03} or other morpho-photometric characteristics and environments \citep{Trager00, Trager06, Trager08}. 
For this reason, we need to combine the UVB and VIS arms, in order to simultaneously fit a larger number of stellar absorption features arising from different chemical species (e.g. CaK, CaH, G4300, Mg, Fe, Na, Balmer lines, TiO). 
However, since the two spectra from the two different arms of XSH have different nominal resolutions ($\mathrm{R}_{\mathrm{VIS}}=5000$, $\mathrm{R}_{\mathrm{UVB}}=3200$), a spectral convolution needs to be performed in order to bring the VIS and UVB spectra to the same final resolution, before combining the two sections. 
%However, since the two spectra, coming from the two different arms, have different resolution ($\mathrm{R}_{\mathrm{VIS}}=5000$, $\mathrm{R}_{\mathrm{UVB}}=3200$), before such combination can take place, a spectral convolution has to perform, bringing the VIS and UVB spectra to the same final resolution. 
For this purpose, we first shifted the spectra to the rest-frame wavelength %we first restframe the spectra 
and then convolved each spectrum with a Gaussian function with a variable sigma (following the prescription of \citealt{Cappellari17}).  
The kernel of the Gaussian broadening function is computed as follows: 
\begin{equation}   
   \mathrm{kernel}(\lambda)= \exp \left( \dfrac{-\lambda^{2}}{2\,\eta(\lambda)^{2}}\right)
,\end{equation}   
where $\lambda$ is the wavelength measured in \AA\, and $\eta$ is defined as 

\begin{equation}   
   \eta(\lambda)= \dfrac{\sqrt{\mathrm{FHWM}_{\mathrm{fin}}^{2} - \mathrm{FHWM}_{\mathrm{XSH}}(\lambda)^{2}}}{2.3548}
,\end{equation}   
with a full width half maximum  %\LEt{reminder about note 3}
FWHM$_{\mathrm{XSH}} =\lambda/\mathrm{R}_{\mathrm{arm}}$ for each arm, and FWHM$_{\mathrm{fin}} = 2.51$\AA\, at the restframe wavelength. This latter final value was chosen to be equal to that of the stellar templates and the single stellar population models used in the next section. 
%to minimize template mismatch during the pPXF fitting described below.
The final combined 1D spectra are plotted in Figure~\ref{fig:final_1dspec} in units of $10^{-17}$ erg s$^{-1}$cm$^{-2}$\AA$^{-1}$ and at rest-frame wavelength. 
In the figure, red dashed lines highlight the stellar absorption lines, typical of ETGs, which are used to derive stellar kinematics and population parameters in the next sections. 
Grey vertical shaded boxes show the regions where the UVB and VIS spectra were joined, while purple ones show the region contaminated by telluric lines that we corrected with \mbox{{\tt MOLECFIT}}. To avoid possible problems of flux calibration at the boundaries of the two arms, we manually assigned a very large error to these pixels so that they would not be considered in the stellar template fitting procedure described in the next sections. %Since some residuals are still visible, we assign to these pixels larger errors and thus a smaller weights during the fit. 

We infer the final S/N in the central part of the final spectra, from a region of 500 \AA, centred around the Mg$_{b}$ and Fe5270, Fe5335 absorption lines (5000-5500 \AA\, rest frame). The resulting values are reported in Table~\ref{table:kinematics}. We note that, for two of the objects, the resulting S/Ns are slightly lower than expected. For the next semester of XSH observations (P105), we revised the integration times on the basis of the results obtained from these three spectra.

\subsection{Stellar kinematics}
\label{sec:kinematics}
The stellar velocity and velocity dispersion were derived using the Penalised Pixel-fitting software (\ppxf; \citealt{Cappellari04,Cappellari17}). 
In particular, we used the latest python version of \ppxf, v7.2.0.\footnote{The package is available here: \href{https://pypi.org/project/ppxf/}{https://pypi.org/project/ppxf/}.}

As suggested by the author of the code, to fit the kinematics we used %a multiplicative Legendre polynomial of 12 grade and 
an additive Legendre polynomial but did not use any multiplicative ones. This approach helps to correct the continuum shape during the fit, which is only performed to obtain the two first moments of the Gauss-Hermite series (velocity and velocity dispersion). The final chosen grade for the polynomial additive degree is 20, but we investigate the impact of a different choice in the Appendix~\ref{app:polinom}. 

The fit was performed between $3500$-$6500$ \AA, where strong stellar absorption lines are present (CaK, CaH, H$\gamma$, G-band, H$\beta$, Fe, Ca, Mg, NaD, see  Fig.~\ref{fig:final_1dspec}).  
During the fit, we carefully set the good-pixel region by hand, manually masking out the region where the UVB and VIS spectra were joined (i.e. the grey regions in Fig.~\ref{fig:final_1dspec}) and assigning larger errors to the regions where the deepest telluric line in the optical range was removed by \mbox{{\tt MOLECFIT}} (i.e. purple regions in Fig.~\ref{fig:final_1dspec}).  
In addition, we also used a sigma clipping method similar to that described in \citet{Cappellari+02}.  
In practice, we ran the fit for the first time and selected all the galaxy pixels that deviated more than 3 sigma, in terms of the  %RMS%\LEt{doesn't need an intro :)}
 RMS, \footnote{We note that for the fit we do not use the noise provided by the pipeline but an RMS-noise-computed smoothing the spectra with a medial filter with a kernel of 13 pixels.} %\LEt{Is there a way we can say this accurately without saying kernel twice?}.}, 
 from the best fitting template spectrum. We then repeated the fit a second time with the same parameters as before but masking out all the pixels selected during the previous step.  
%The upper panel in Figures~\ref{fig:SP_resultsJ0224}, \ref{fig:SP_resultsJ0314} and \ref{fig:SP_resultsJ0847} show 

To fit the LOSVD, we did not apply any linear regularisation to the weights of the stellar templates (i.e. we set the corresponding keyword in \ppxf\, REGUL=0), which are instead tested to recover the stellar populations, as described in Sect.~\ref{sec:population}, and, in more detail, in Appendix~\ref{app:regul}. 

As stellar templates for fitting the velocity and velocity dispersion, we tested both single empirical stars and single stellar population (SSP) models. We thus verified that the kinematical results do not depend on the template sets used for the fit. 
In particular, as empirical stars, 
we employed the full MILES stellar library, which consists of 985 stars spanning a large range in atmospheric parameters (\citealt{Sanchez-Blazquez+06}, \citealt{Falcon11}). The spectra cover the 3525-7500 \AA\, range and have a spectral resolution of FWHM = 2.51\AA.\footnote{The MILES library, v9.1, is available on the project's website: \href{http://research.iac.es/proyecto/miles/pages/stellar-libraries/miles-library.php}{http://research.iac.es/proyecto/miles/pages/stellar-libraries/miles-library.php}.} 
As SSPs, we instead used the MILES models by \citet{Vazdekis15} with a bimodal stellar IMF with a fixed slope of $\Gamma = 1.3$ and BaSTI theoretical isochrones.\footnote{
\href{http://www.oa-teramo.inaf.it/BASTI}{http://www.oa-teramo.inaf.it/BASTI}.} The stellar population parameters of the models, which are also used to study the stellar populations of the three galaxies, are described in more detail in the next section. 

The result of the kinematics fit obtained with the SSP models is listed, for each galaxy, in Table~\ref{table:kinematics}. We  note that, since the spectra are in the rest-frame wavelength, the velocity that we quote, which is the systemic velocity relative to the nominal redshift of the system, should be consistent with zero. In the same table, we also list the S/N calculated around the Mg$_{b}$, summing up the fluxes between 5000 and 5500 \AA and dividing them by the sum of the errors over the same range and the reduced $\chi^2$ of the fit. 
We do not list the results obtained with stars, but these are always consistent within a 1-sigma error with those obtained using SSP models. 
 
\begin{figure}
    \centering
    \includegraphics[width=\columnwidth]{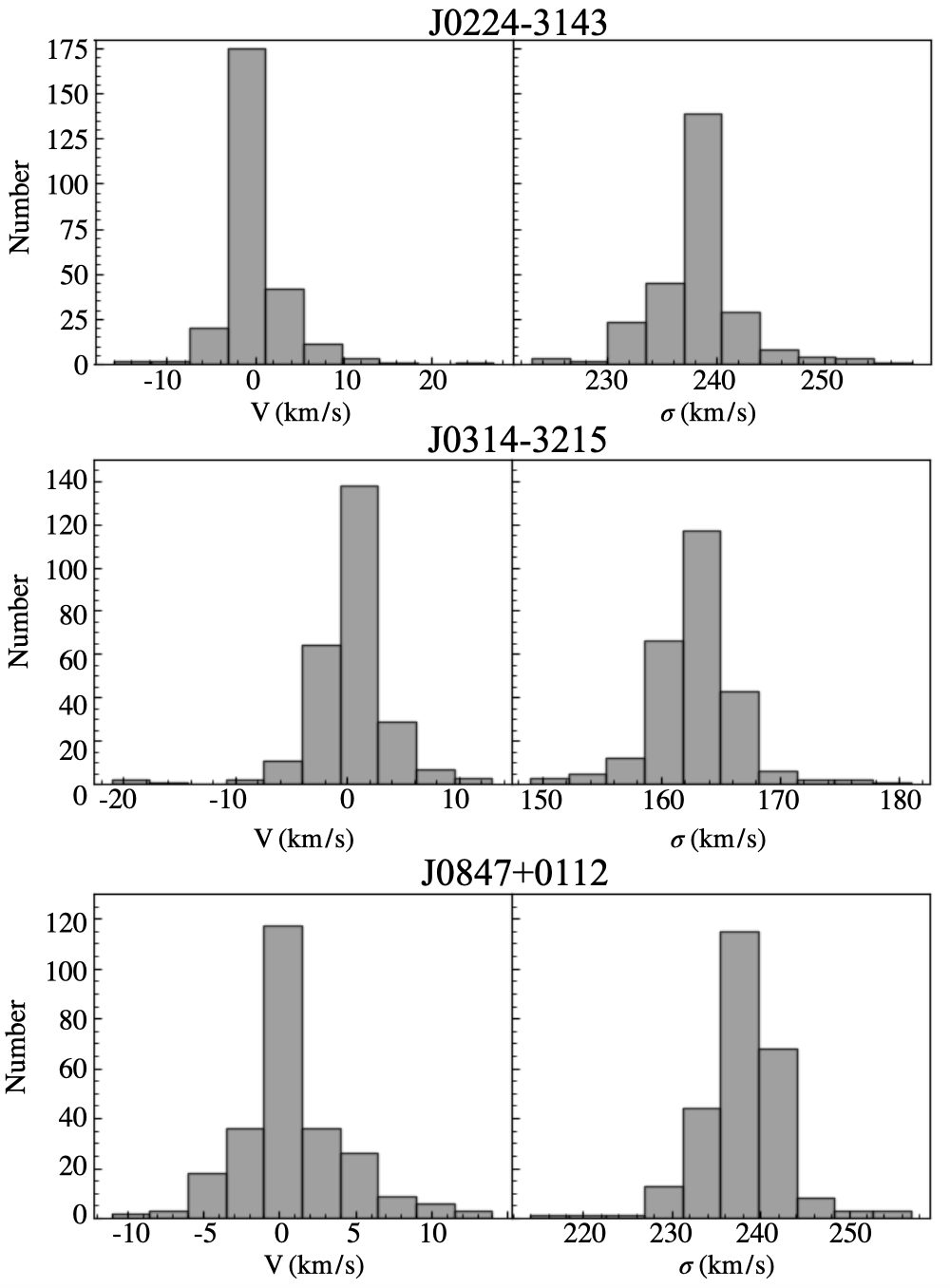}
    \caption{Distribution of velocities and velocity dispersions computed repeating the kinematical fit 256 times, changing the parameters and setup, as described in the text, to estimate the uncertainties associated with velocity and velocity dispersion. The values reported in Table~\ref{table:kinematics} are instead those obtained with the best-fit configuration on the original observed spectrum.} 
    \label{fig:histo_kinem}
\end{figure}

Uncertainties on the estimated values for the LOS kinematics due to: i) a variation of the additive polynomial; ii) the use of different templates; and iii) re-scaling of the noise are taken into account, on top of the random errors given in output by a single fit. % and the systematics uncertainties that we estimated in the following way. 
For this purpose, we developed a semi-automatic routine that repeats the fit multiple times, following a classical Monte Carlo perturbation. In particular, the routine simulates the observed spectrum 256 times, pixel by pixel, each time adding a random noise generated using a Gaussian distribution with the observed flux at that pixel as central value and the RMS noise as standard deviation. 
Then, we ran the fit on these 256 new spectra, varied 
the grade of the additive polynomial between 16 and 24 ($\pm 4$ from the optimal value), and rescaled the noise by a constant factor in order to maintain the reduced $\chi'^{2}$ between 0.9 and 1.1.\footnote{The $\chi'^{2}$ is equal to the $\chi^{2}$ divided by the degree of freedom, which is equivalent, in this case, to the number of good pixels, N$_{\mathrm{good\,pixels.}}$}  %These are, to our knowledge, the main sources of known systematics.   
Hence, in this way, we obtained a distribution of velocity and velocity dispersion measurements with associated uncertainties, which are shown in Figure~\ref{fig:histo_kinem}. The final quoted values of the uncertainties listed in Table~\ref{table:kinematics} were then computed as standard deviations. The central values that we report are instead the values obtained from the original observed spectrum using the best-fit configuration (minimum $\chi^{2}$, ADEGREE=20).

\begin{table*}
\caption{Final results of the kinematics \ppxf\, fit obtained with ADEGREE = 20 and MDEGREE =-1 in the spectra range between $3500$-$6500$ \AA\, and a comparison with the results presented in S20 (last five columns, separated by a vertical line). We note that the S/N is calculated around the Mg$_{b}$ for the XSH spectra, in a bluer region in S20 and over a larger wavelength window ([4500-6500]\AA) for the GAMA spectrum of KiDS J0847+0112. The velocity dispersion of the GAMA spectrum is computed using \ppxf\, in the same configuration used for the XSH spectra.  %The physical aperture corresponding to the 5 pixels over which the spectra have been extracted is also reported in the table, in unit of the effective radius (R$/$\Reff). 
%The last two columns can be directly compared because in both cases the velocity dispersion is computed at the \Reff. 
The uncertainties on the XSH velocity and velocity dispersion values are computed as standard deviation of the distributions shown in Figure~\ref{fig:histo_kinem}, whereas the central values are obtained with the best configuration.}. 

\label{table:kinematics} 
\centering 
\begin{tabular}{ccccc|ccccc} 
\hline
\hline
KiDS & V$^{*}$ & $\sigma_{\star}$ & $\chi'^{2}$ & SNR  & ID S20 &  SNR$_{\mathrm{S20}}$ & SNR$_{\mathrm{GAMA}}$ &  $\sigma_{\star,\mathrm{S20}}$ & $\sigma_{\star,\mathrm{GAMA}}$ \\
%R$/$\Reff & ID S20 &  $\sigma_{\mathrm{S20}}$ \\ %& $\sigma (\Reff)$\\    
& (\kms)  & (\kms) & & (5000-5500\AA) &  & (3600-4600\AA) & (4500-6500\AA) &  (\kms) &  (\kms) \\
\hline                      
J0224-3143 & $-2\pm 4$ & $259\pm 4$ & 0.94 & 31 %& 1.25 
& 41 & 3.69 & & $211\pm 40$ &  \\ %& $263\pm 4$\\ 
J0314-3215 & $6\pm 4$ & $183\pm 4$ & 1.00 & 19 %& 2.5 
& 45  & 3.09 & & $195\pm 46$ &  \\ %& $195\pm 4$ \\ 
J0847+0112 & $-23\pm 4$ & $250\pm 5$ & 0.98 & 15 %& 0.85 
& L4 &   &  15.56 & &  $276\pm7$\\ %$259\pm 75$  \\ 
% & $247\pm 5$\\ 
\hline        %inserts single line
\hline
\end{tabular}
\end{table*}

An estimate of the stellar velocity dispersion at the effective radius for two of these three objects %J0224-3143 and J0314-3215 
was already presented in S20. 
%\textcolor{red}{TO BE DECIDED: Do we apply the aperture correction? It will not change things drammatically anyway!}
%In order to fairly compare our results with the previously published ones, we need to correct for the different apertures. To this purpose, we use the formula given in Eq.~1 of \citep{Cappellari+06} to convert our measurements to the \Reff. 
Given the much lower S/N of the spectra used in that paper, the resulting uncertainties on the values were, of course, much larger than the ones inferred here.  Nevertheless, the values are very close to each other for KiDS J0314-3215, and within 1.5$\sigma$ error for KiDS J0224-3143.  KiDS J0847+0112 has been targeted by the GAMA Survey. We downloaded the GAMA spectrum and calculated the velocity dispersion using \ppxf\, with the same configuration. Also in this case, the result is in fair agreement and consistent within 2$\sigma$ errors. The values of the velocity dispersion presented in S20 (and GAMA) from the low-quality spectra are listed in  Table~\ref{table:kinematics}, along with the S/N and the ID of the objects. %and those calculated using the aperture correction are all listed in Table~\ref{table:kinematics}. }

%\subsection{[$\alpha/$Fe] abundances with line-indices measurements}

\subsection{Stellar populations}
\label{sec:population}
As already stated, our working hypothesis is that relics are galaxies that have been formed via a short and intense star formation `burst' during the first phase of the formation scenario proposed for massive ETGs, and then have evolved passively and undisturbed ever since. 
As such, they should ideally not have any signs of recent star formation episodes and be well represented by a single, very old stellar population with super-solar $\alpha$ abundance and metallicity.  
The large [$\alpha$/Fe] originates because the quenching of the star formation occurs before Type Ia supernova explosions can pollute the interstellar medium with iron \citep[e.g.][]{Gallazzi+06, Gallazzi+14,Thomas+05}.

The super-solar metallicity is expected, because these objects are relatively massive, and thus, following the mass-metallicity relation \citep[e.g.,][]{Gallazzi+05}, one would expect that they are populated mainly by metal-rich stellar 
populations. 
%\footnote{We note that we could have extended the \ppxf fitting to a third dimention, however, measuring the [$\alpha$/Fe] from indices and from full spectral fitting might not be the same. In fact, from line indices, one is mainly sensitive to [Mg/Fe] since  the ratio is derived from from the absorption line indices Mg$b$ and $\langle$Fe$\rangle$ (e.g. \citealt{2005ApJ...621..673T}). With full spectral fitting, instead, ...}.  

\begin{figure}
    \centering
    \includegraphics[width=9cm]{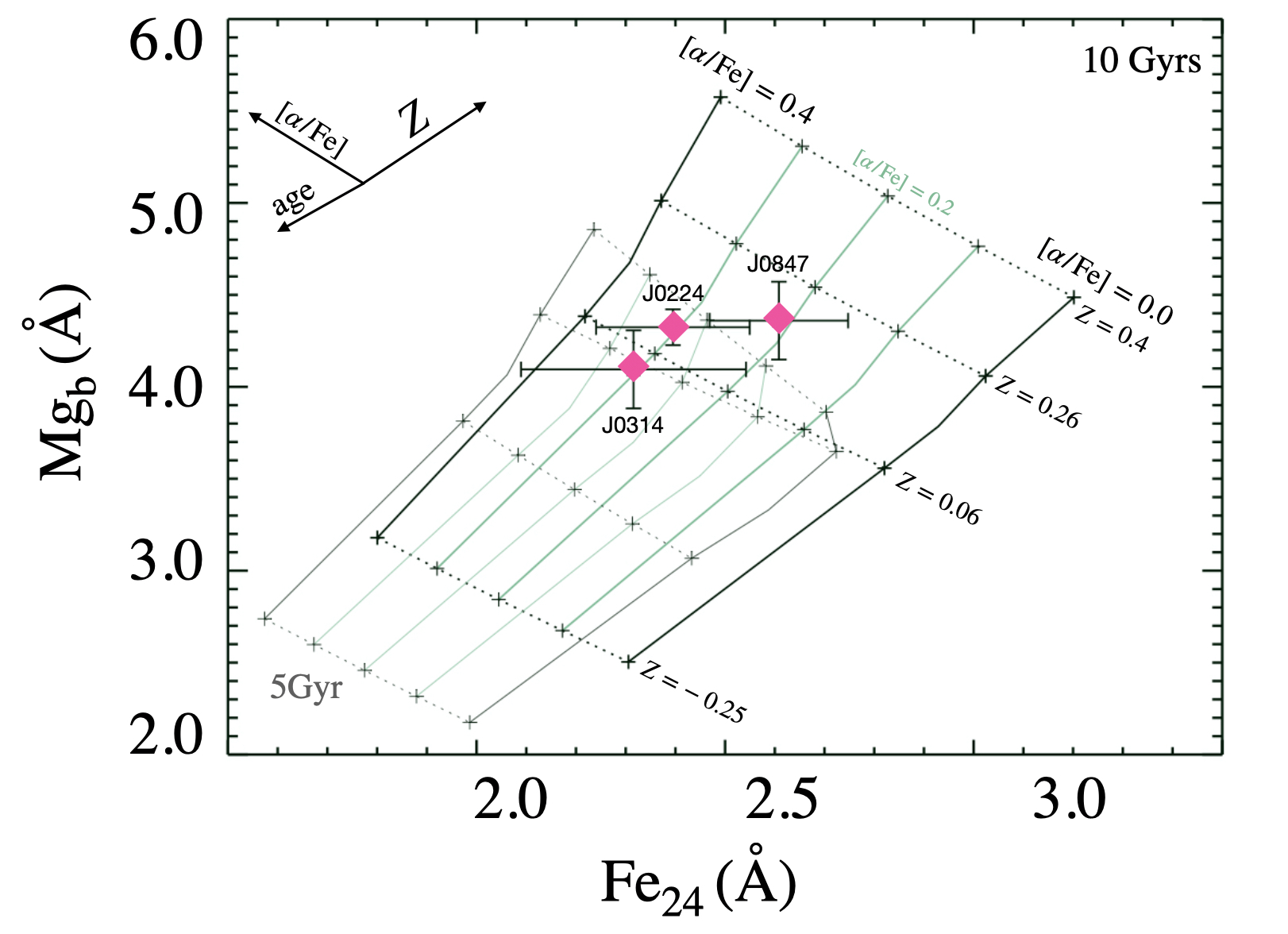}
    \caption{Line-index inference on the [$ \alpha$/Fe] abundance of the relic candidates. In this index-index plot, the classical Mg$_b$ index \citep{Trager08} is plotted against the Fe$_{24}$ index, which is the mean of the 24 iron lines present in the spectral range we considered here ($3500-6500$). The grid represents the strengths of MILES SSP models with varying metallicity (dotted) and [$\alpha$/Fe] (solid; we colour-coded -in black- the original models and -in green- the interpolated ones), as indicated in the plot but with a fixed age of 10 Gyr. To show the effect of a varying age, which does not affect the inference on [$\alpha$/Fe], we also plot younger models with an age of 5 Gyr with less thickness. Black arrows in the top left of each panel indicate the direction in which the model grid will move if the corresponding parameters increase. Both galaxies and SSP line strengths are calculated after convolving the spectra to a final, common resolution of $\sigma=270$ \kms.}
    \label{fig:alphafe}
\end{figure}

Furthermore, \ppxf\, allows one to extract distributions only in the 2D parameter
space
defined by age and metallicity, thus, for an estimate of [$\alpha$/Fe], we
took a different approach. We first inferred the [$\alpha$/Fe] via line-index measurements and then ran the full-spectral fitting code using only models with the `right' value of [$\alpha$/Fe] for each system. 
In detail, we used the SPINDEX code from \citet{Trager08} to calculate the line-index strengths of the Mg$_b$ and the many iron (Fe) lines present in the wavelength range at our disposal. 
We then produced the classical Mg$_b$-Fe index-index plot, which allowed us to obtain an estimate of the [$\alpha$/Fe] and a first guess of the total metallicity for the three objects. To minimize the error bars and obtain a constraint that was as secure as possible, and as independent as possible on other elemental abundances (e.g. [Ti/Fe], as for Fe5015), we took the mean of 24 different iron lines,\footnote{Fe3619, Fe3631, Fe3646, Fe3683, Fe3706, Fe3741, H10Fe, FeBand, Fe4033, Fe4046, Fe4064, Fe4326, Fe4383, Fe4457, Fe4592, FeII4550, Fe4920, Fe5015, Fe5270, Fe5335, Fe5406, Fe5709, Fe5782, and Fe6497. The band-pass definition and reference for all the indices can be found on the MILES SSP website (\url{http://research.iac.es/proyecto/miles/}).} creating the Fe$_{24}$ index that is plotted on the x-axis in  Figure~\ref{fig:alphafe}. 
We overplotted the line strengths obtained from MILES SSP models with varying total metallicity and [$\alpha/$Fe] and two different fixed values of the age (10 and 5 Gyr). Both galaxies and SSP spectra were smoothed to a common resolution of $\sigma = 270$ \kms \, before the indices were computed. %We note that using instead the classical and more broadly used <Fe>=Fe5270+Fe5335  

Given that the publicly available version of the MILES models only allows for two different [$\alpha$/Fe] abundances (0.0 and 0.4),  we linearly interpolated the models between these values (for each different age and metallicity values) to build a finer grid ($\Delta=0.1$). The interpolated models are plotted in green, while the original ones are plotted in black. We simply assumed the one of the closest models
as the [$\alpha$/Fe] value and assigned an error bar equal to the step between the models ($0.1$).
%(Fe3619, Fe4033, H10Fe, Fe5015 and Fe5782)

%Fe_J0224 = median([Fe4920(0),Fe5270(0),Fe5335(0),Fe6497(0)])
%Fe_J0314 = median([FeBand(1),Fe4033(1),Fe4046(1),Fe4064(1),Fe4383(1),Fe4457(1),Fe4592(1),Fe5015(1),Fe5335(1),Fe5782(1)])
%Fe_J0847 = median([FeBand(2),Fe4046(2),Fe4064(2),Fe4383(2),Fe4457(2),Fe4531(2),Fe4592(2),Fe4920(2),Fe5015(2),Fe5270(2),Fe5335(2),Fe5782(2)])

By making use of \ppxf, we constrained, at this point, the mass-weighted stellar age and metallicity for the three relic candidates, after fixing the [$\alpha$/Fe] to the value read from the index-index plot. We ran the fit using only SSP models with [$\alpha$/Fe] = 0.2 for KiDS J0847+0112 and of 0.3 for the other two systems as templates, but, given the error bars from Figure~\ref{fig:alphafe}, we tested the impact that changing the [$\alpha$/Fe] has on the kinematics and on the stellar populations in Appendix~\ref{app:alpha}. 

This time, again following the prescriptions given within the code documentation, we remove the additive polynomial (ADEGREE=-1), which could affect the line strength of the spectral features, and use a multiplicative one instead (MDEGREE=10, but we tested the impact of this assumption on the results in Appendix~\ref{app:polinom}). 

\begin{figure*}
 %   \centering
    \includegraphics[width=19cm]{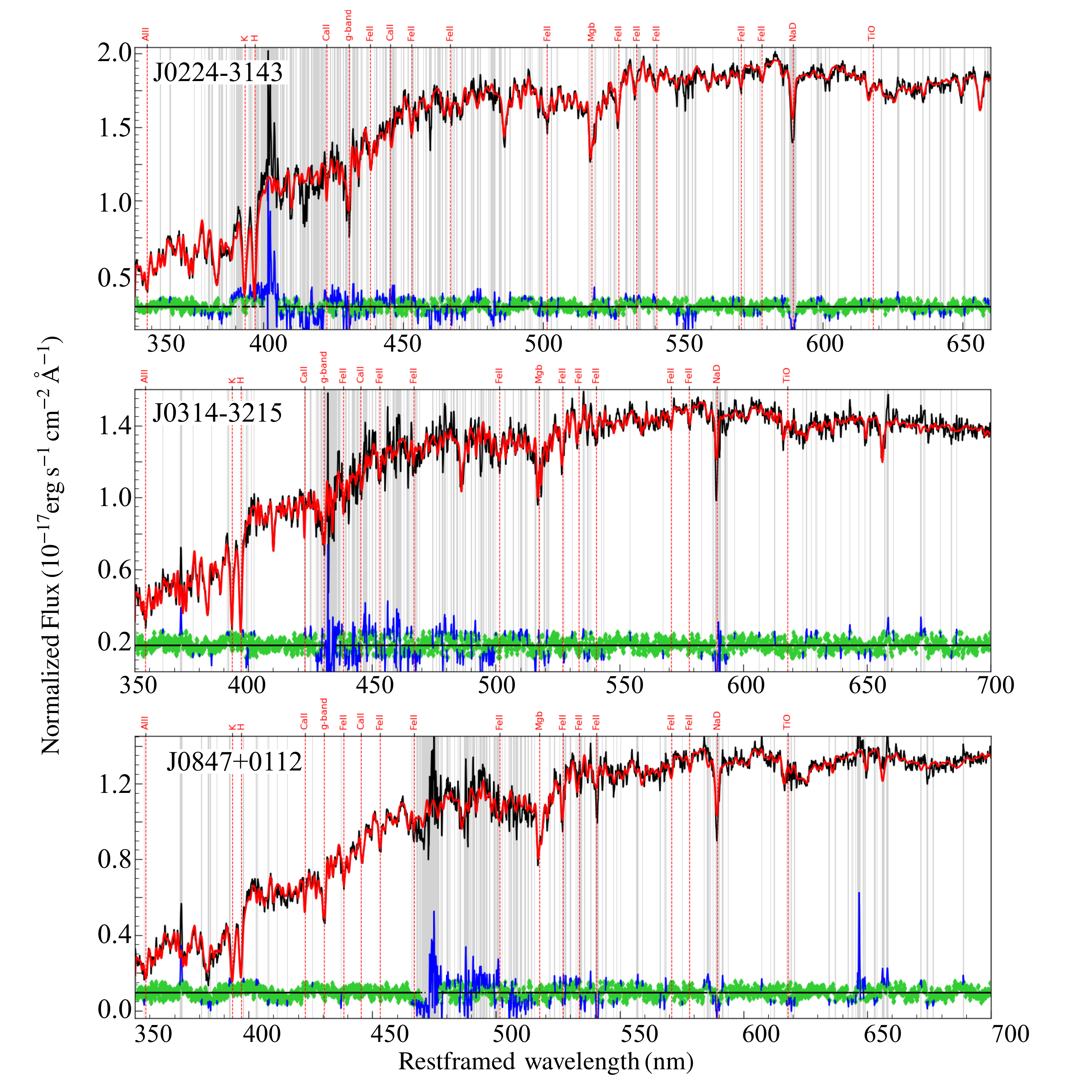}
    \caption{Final results of stellar population \ppxf\, fits on KiDS J0224-3143 (top row), KiDS J0314-3215 (middle) and KiDS J0847+0112 (bottom). Each panel shows the galaxy spectrum (black) with the best-fit stellar template (red) overplotted on it. The residuals for the good pixels are shown in green; blue lines are the pixels that have been masked out via our sigma-clipping routine, while grey shaded regions have been excluded from the fit by hand. }
    \label{fig:ssp_plots}
\end{figure*}

As starting guess for velocity and velocity dispersion, we use the values obtained in Sec.~ \ref{sec:kinematics},  which are also listed in Tab.~\ref{table:kinematics}. The fitting results %and the distribution of the weights obtained with the MAX\_REGUL 
for KiDS J0224-3143, KiDS J0314-3215, and KiDS J0847+0112 are plotted in Figure~\ref{fig:ssp_plots}.  
The galaxy 1D spectrum is plotted (in rest-frame wavelength) in black, and the best-fit template is overplotted in red. The residuals of the fit are shown in green; blue lines are the pixels that have been masked out via our sigma-clipping routine, while grey shaded regions have been excluded from the fit by hand because they correspond to tabulated emission or sky lines (as highlighted on the top x-axis of the panel). 

We note that the velocity dispersion values obtained with this new \ppxf\, setting are  slightly lower, but overall consistent (within 2$\sigma$ uncertainties), with those given in the previous section (given in Table~\ref{table:kinematics}).  The impact of performing the fit with $\alpha$-enhanced SSP models is described in the Appendix~\ref{app:alpha}.

\begin{figure}
    \centering
    \includegraphics[width=8cm]{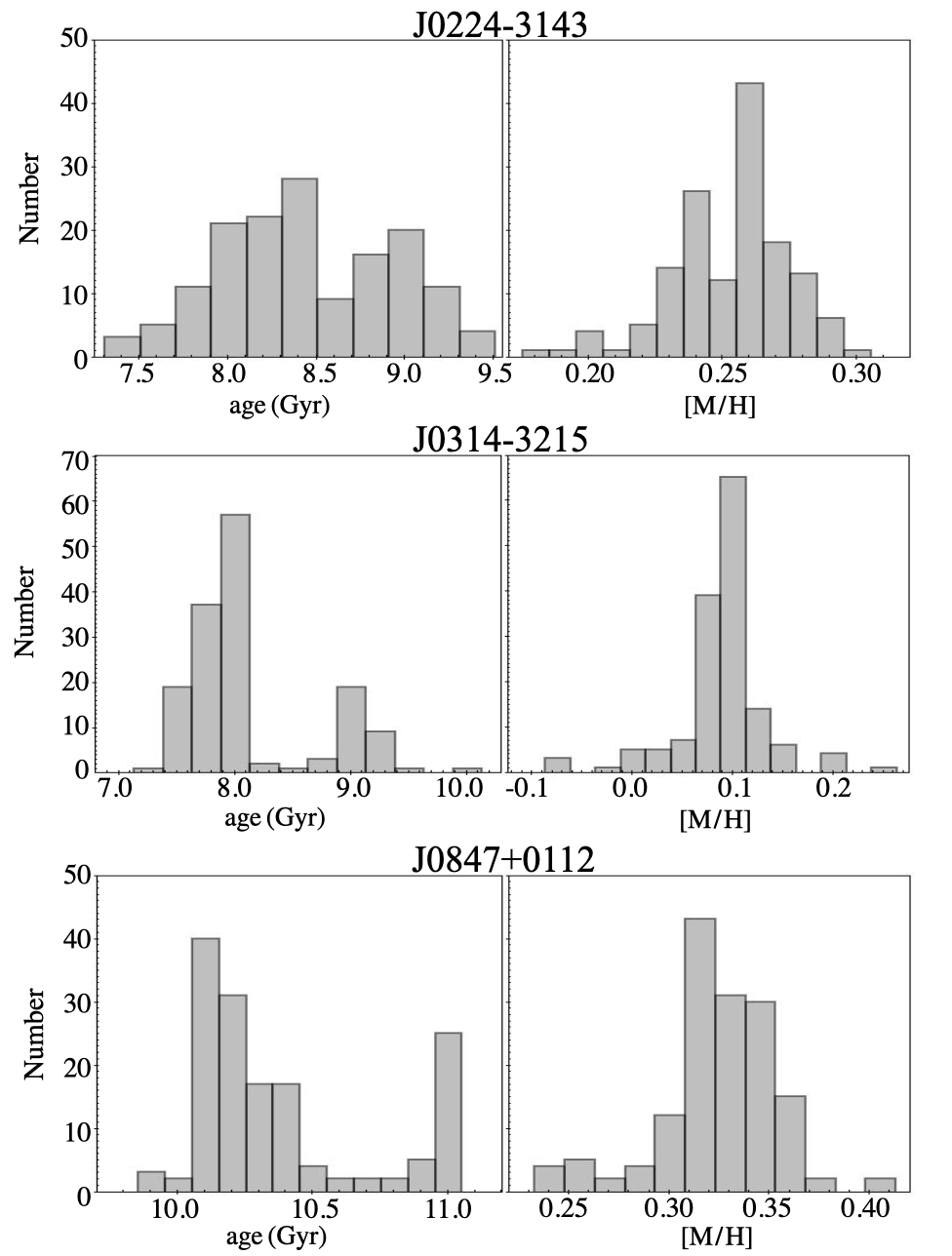}
    \caption{Histograms of the distribution of mass-weighted mean ages and metallicities inferred when repeating the fit 256 times; changing the parameters, the set-up, and the regularization, as described in the text. These distributions were only used to estimate the uncertainties on the values reported in Table~\ref{table:ssp}. The mean ages ([M/H]) of the distributions shown here are not identical to the values found with the best possible setup (listed in the table) but do not differ by more than 1 Gyr (0.04 dex).} %These are given in the table, but do not differ by more than 1 Gyr (0.04 dex).}
    \label{fig:histo_popul}
\end{figure}

Furthermore, using the same semi-automatic routine already employed to estimate the errors on the kinematic parameters, we also obtain a realistic estimate of the standard deviation on the stellar population parameters. We repeated the fit 256 additional times on each spectrum (i.e. for each system) at every run, randomly simulating a new spectrum according to the $1\sigma$ RMS noise level obtained during the first fit, changing the MDEGREE from 6 to 14 ($\pm4$ from the chosen value) and the regularization from 0 to MAX\_REGUL. %, and the fitted region. 
The distribution of the different meaningful values of the age and [M/H] for each galaxy are plotted in Figure~\ref{fig:histo_popul}. The standard deviation of each distribution is assumed as uncertainties on the values listed in Table~\ref{table:ssp}. 
We note that we only used this procedure to compute the standard deviation on the central values of the mass-weighted stellar population parameters, which were instead calculated in the best possible configuration (producing the best $\chi'^{2}$ and with MDEGREE=10). 

We performed the stellar population fit in two different ways: once fixing the regularisation parameter to null (REGUL = 0), and a second time considering instead the maximum regularisation still consistent with the galaxy spectrum and its S/N (MAX\_REGUL). Hence, with MAX\_REGUL, the derived star formation history (SFH) is %will be\LEt{will be or is? is it in the future. Reminder to please check your tenses carefully} 
the smoothest (minimum curvature or minimum variation) one %, which is 
still consistent with the spectral observations. 
On the other hand, with REGUL = 0, the derived SFH gives a much larger weight to a much smaller number of  %weighs much more a smaller number of 
templates, %\LEt{\textless-- not sure what this first part of the sentence means}, 
thus possibly discarding secondary star formation episodes.  
%Whereas with REGUL = 0, the derived SFH will only include the major star formation episodes that each galaxy experienced.  
Considering these two extreme cases, is therefore important to interpret the inferred SFH and draw conclusions on the `relicness' of the three galaxies, which is discussed in Section~\ref{sec:results}. 
Finally, in Appendix~\ref{app:regul}, we describe tests on more values for the REGUL parameter and study the impact that this has on the derived stellar population parameters and on the SFH. 
To set the MAX\_REGUL, we used a total number of SSP models ranging from 90 to 110, %126 to 154 
depending on the galaxy, with ages ranging from 1 to the age of the Universe at the redshift of each object (9.5, 10.5, and 11.5 for KiDS J0224-3143, J0314-3215, and J0847+0112, respectively) with steps of $\Delta t = 0.5$ Gyr and five %seven 
different total metallicity bins (in terms of the star's total metal abundance, %[M/H] = $\{-0.66, -0.35, -0.25,+0.06,+0.15,+0.26,+0.40\}$. 
 [M/H] = $\{-0.66,-0.25,+0.06,+0.26,+0.04\}$. 
 For each galaxy, we only used models with the [$\alpha$/Fe] value we estimated from Figure~\ref{fig:alphafe}. The MAX\_REGUL found for each galaxy is reported in Table~\ref{table:ssp}. We note that the spectrum of KiDS J0314-3215 allows for a much smoother solution (MAX\_REGUL=600) and thus requires a more extended SFH, with respect to the other two objects. We comment further on this in the next section.  

\section{Results}
\label{sec:results}
The final results of the \ppxf\, stellar population fitting are reported in Table~\ref{table:ssp}, both for the unregularised (REGUL = 0) case and that with the maximum regularisation (REGUL = MAX\_REGUL). In the table, for each system we list the mass-weighted stellar population parameters with associated uncertainties and also the formation redshifts (z$_{\mathrm{form}}$) for both cases. Formation redshifts were obtained with  %general cosmology, 
$\Lambda$CDM general cosmology in the following way. First, we calculated the age of the Universe at the redshift of each galaxy, $t_{\mathrm{uni}}(z_{\mathrm{gal}})$, then we subtracted the age of the Universe at the formation redshift, $t_{\mathrm{uni}}(z_{\mathrm{form}})$, and made it equal to the mass-weighted age estimated from the SSP fitting (also listed in the table): 
\begin{equation}
 t_{\mathrm{uni}}(z_{\mathrm{gal}})-t_{\mathrm{uni}}(z_{\mathrm{form}}) = t_{\mathrm{SSP}}
.\end{equation} 
%(also listed in the table, which for short SFH as the ones obtained here can be approximated to {\bf the age of the formation}). 
Finally, we solve for z$_{\mathrm{form}}$.

\begin{table*}
\caption{Results of the stellar population \ppxf\, fit obtained for each galaxy with MDEGREE = 10 and ADEGREE =-1 in the spectra range from $3500$-$6500$ \AA, for REGUL = 0 and for REGUL = MAX\_REGUL (listed in the table for each system). The uncertainties associated with the measurements include both the random error obtained directly from the fit and the systematics, estimated repeating the fit multiple times, changing the configuration and setup, also including the regularization of the fit (always between 0 and MAX\_REGUL). We also list the age of the Universe at the redshifts of the system, the [$\alpha$/Fe] inferred from the line indices, and the redshift at which these objects formed. %, and during the 256 following fits performed with a more limited set of models. 
See the text (Sect. \ref{sec:population}) and Appendix \ref{app:regul} for more details.}        
\label{table:ssp} 
\centering 
\begin{tabular}{c c c c c c c c } 
\hline
\hline
KiDS & REGUL & Age  & [M/H]  & $\chi'^{2}$ & Age of the & [$\alpha$/Fe]$^{\star}$   & z$_{\mathrm{form}}$ \\
&   &  (Gyr) & (dex) & & Universe (Gyr) & (dex) &   \\
\hline    
\hline
J0224-3143 &  0  &  $9.1\pm0.4$ & $0.24\pm0.03$ & 0.9920 & 9.5 & $0.3\pm0.1$  &  10.65  \\ 
J0224-3143 &  11 &  $8.6\pm0.4$ & $0.24\pm0.03$ &  1.0297 & 9.5 &  $0.3\pm0.1$  & 6.02   \\ 
\hline
J0314-3215 & 0  &  $9.1\pm0.5$  & $0.10\pm0.04$ &  1.0000 & 10.4 &  $0.3\pm0.1$  & 4.65 \\ 
J0314-3215 &  600  &  $7.9\pm0.5$  & $0.10\pm0.04$ & 1.0267 & 10.4 &   $0.3\pm0.1$  & 2.64 \\ 
\hline
J0847+0112 &  0  &  $11.0\pm0.3$ &  $0.35\pm0.03$ & 1.0001 &  11.5 &  $0.2\pm0.1$  & 9.20  \\
J0847+0112 &  62  &  $10.6\pm0.3$ &  $0.34\pm0.03$ & 1.0264 & 11.5 &   $0.2\pm0.1$  & 6.04  \\
\hline
\hline       
\end{tabular}
\end{table*}

The stellar population parameters derived for the three relic candidates analysed here confirm that these objects are indeed mainly populated by old stars. This is also fully consistent with our previous findings,  based on photometric colours (see Fig.~\ref{fig:old_ages_color}) and low-resolution spectroscopy (T18, S20).\footnote{We stress that the spectra previously available for two of these three systems, which showed no signs of emission lines and were typical of an evolved population, had a very low S/N ($\sim3$ per \AA), which prevented us from performing a detailed stellar population analysis. For KiDS J0847+0112, the results obtained from the GAMA spectra, of comparable S/N but lower spectral resolution (R$\sim1300$), are in good agreement with those obtained from the new XSH spectrum.}  

\begin{figure}
    \centering
    %\vspace*{-10mm}
    \includegraphics[width=\columnwidth]{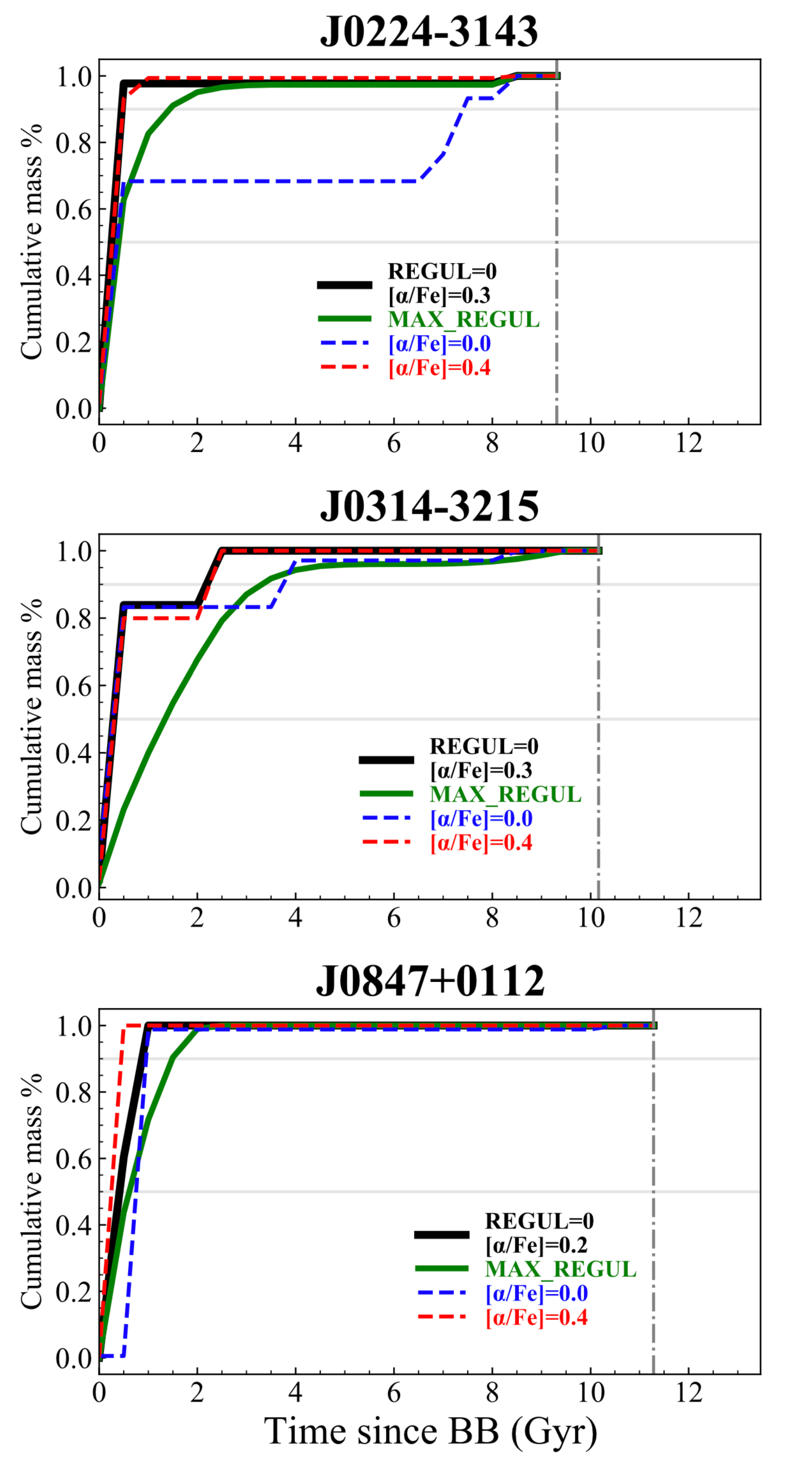}
    %\vspace*{-10mm}
    \caption{Star formation histories obtained with \ppxf\, from the spectra of the three objects extracted within an aperture of radius $r=0.4\arcsec$, corresponding to five XSH pixels. The time x-axis corresponds to the time since the BB, and the vertical solid line in each panel shows the age of the Universe at the redshift of the corresponding object. Black lines show the cumulative stellar mass fraction inferred from the unregularised mass-weighted \ppxf\, fit with the best value of [$\alpha$/Fe] estimated from line indices (Fig.~\ref{fig:alphafe}).  Green lines instead show the results of the fit performed with the regularization set to the MAX\_REGUL. Finally, coloured dashed lines show the impact of performing the fit (with REGUL = 0) using models with different [$\alpha$/Fe]. } 
    %, and then a second one occurred rougly after 5.5 Gyrs from the BB.    %J0314-3215 (middle panel) }
    \label{fig:sfh}
\end{figure}

In order to assess whether there systems are relics or not, in Figure~\ref{fig:sfh}, we plot the cumulative stellar mass formed at each time, starting from the Big Bang (BB). We do so for the main setups and assumptions we made in the paper (i.e. for the two extreme regularisation values, assuming the [$\alpha$/Fe] as computed via line-index analysis or taking the two extreme [$\alpha$/Fe] values permitted by the SSP models.

From the figure, it is clear that KiDS J0847+0112 (bottom panel) has the most `extreme' star formation history, when considering all the possible sources of uncertainties due to the different configurations and setups. It formed the totality of its stellar mass at most 2 Gyr after the BB, most likely even before and over very short timescales ($\sim0.5$-$1.0$ Gyr), and then it had no further star formation episodes.  %This result stands for all the considered cases. 

The star formation history of KiDS J0224-3143 (top panel) is not that different from that of KiDS J0847+0112 for the best configuration. %It depends even less on the regularization.  
However, it is fair to highlight that considering [$\alpha$/Fe]$=0$, which is however %though \LEt{not sure what you mean by though here. And what exactly is unlikely?} 
very unlikely given the estimate obtained from the index-index plot, only $\sim70$\% of the stars were assembled during the first formation episode.  

Finally, for KiDS J0314-3215 (middle panel), in the case of an unregularised fit, one main star formation episode formed 80\% of the stars $\le0.5$ Gyr after the BB, and then the assembly was completed with another fast (1 Gyr long) episode that started 2 Gyr after the BB. Considering the unlikely configuration with [$\alpha$/Fe]$=0$, the second star formation burst occurred instead only 4 Gyr after the BB. However, this galaxy has the highest MAX\_REGUL (600), and thus a much smoother SFH is compatible with its spectrum too. In the MAX\_REGUL case, the galaxy completed its assembly (i.e. $\ge 95$\% of the stellar mass in place) only $\sim5$ Gyr after the BB, and therefore we cannot classify it as a relic. The formation redshift of this object is also lower than that of the other two, in both the unregularised ($z_{\mathrm{form, J0314}} = 4.65$) and the regularised ($z_{\mathrm{form, J0314}} = 2.64$) configurations. However, we note that the majority of the stellar mass of this galaxy (80\%) is characterised by an old and metal rich population. Thus, we speculate that although this object cannot be classified as a relic (i.e. having only one short star formation burst early on in cosmic time), it did not go through a numerous series of interactions during the second phase of its formation scenario. This also explains why it is so compact. Finally, for KiDS J0314-3215, we derived a velocity dispersion value lower than that inferred for the other two objects. 
%In addition, for KiDS J0314-3215, we calculate a velocity dispersion value which resulted much lower, although still consistent within the (large) uncertainties, than that inferred in S20 ($\sigma_{\mathrm{S20}}=259\pm75$ \kms). 
We conclude that this system might therefore % have a more extended star formation history and might also 
be less massive than we thought, and thus it might even not classify as a UCMG any longer (not passing the mass threshold). 

Considering the most plausible setups and assumptions, KiDS J0224-3143 and KiDS J0847+0112 %all the possible set-ups and assumptions, KiDS J0224-3143 and KiDS J0847+0112 
are characterised, for at least 95\% in mass, by stellar populations that are only 1-2 Gyr younger than the age of the Universe at their redshifts (at $z=0.3841$, $t_{\mathrm{Uni}}\sim9.5$ Gyr and at $z=0.1764$, $t_{\mathrm{Uni}}\sim11.5$ Gyr). 
They formed through only one star formation burst at very high redshifts ($z_{\mathrm{form, J0224}}=10.65$; $z_{\mathrm{form, J0847}}=9.20$, inferred from the mass-weighted median ages calculated by \ppxf\, for the unregularised case and $z_{\mathrm{form, J0224}}=6.02$, $z_{\mathrm{form, J0847}}=6.04$ for the regularised one) and have a large stellar velocity dispersion, a super-solar metallicity and  super-solar $\alpha$ abundances. KiDS J0314-3215, although also mainly characterised by an old stellar population, has experienced at least two star formation episodes and reached 95\% of the stellar mass $\sim4.5$ Gyr after the BB. 
%In both cases, the best fit is achieved with a MAX\_REGUL<100.

We will %\LEt{in a future paper? If not then please remove 'will'} 
provide a more rigorous definition of 'relicness' when we analyse the full \INSPIRE\, dataset, in future publications. In summary, we can conclude that two out of three galaxies are classified as  relics in the sense that they only had one SF episode, very early on in cosmic time, and they had already assembled 95\% of their total stellar mass 2 Gyr after the BB. 
%This seems also in line with our initial estimate that roughly 2/3 of our sample will be most likely confirmed as `bona-fide' relics, however it is impossible to draw quantitative conclusions with only few objects. }
%However, their SFH is not identical as that of KiDS J0847+0112 is more "extreme", especially when considering all the possible set-ups and uncertainties in the [$\alpha$/Fe] estimate.  The existence of a "degree of relicness", already reported by F17 seem to be confirmed also for these two new objects, as also highlighted in the next section.  

\subsection{Comparison with the three local relics of F17}
Here we briefly qualitatively compare our findings on the first two confirmed relics in the nearby Universe ($0.15<z<0.4$) with the properties of the three local systems confirmed and studied by F17. 
As one can see from Figure~\ref{fig:mass_size}, in a stellar mass-size plane, the \INSPIRE\, UCMGs (green points) and the three local relics of F17 (yellow triangles) lie on the same region, although our objects are on average smaller but slightly less massive. %more compact but less massive. 

In terms of morphology, the two new confirmed relics show a compact disk-like shape (also in line with the axis ratio inferred from \citealt{Roy+18}, $q={0.39,0.38,0.25}$ respectively for KiDS J0224-3143, J0314-3212, and J0847+0112), very similar to that of  massive red nuggets at high-z (z$\sim2$, \citealt{Buitrago+08, vanderWel+11}) and of the local relics. KiDS J0224-3143, however, looks slightly more roundish from the KiDS high spatial resolution $r$-band image shown in Fig.~\ref{fig:cutouts}. This is also the galaxy with the larger S\'ersic index ($n=6.5$), although all three objects have very peaked profiles with $n>4$. This galaxy, according to the stellar mass of $M_{\star}=2.7\times10^{11}M_{\odot}$, estimated in T18, is the most massive relic ever found. 

The velocity dispersion values we inferred for the new nearby relics are lower than the ones reported in F17 for the three local objects. We speculate that this might be due to the fact that, given the smaller apparent sizes on sky of these objects, we are integrating over larger physical radii. If the velocity dispersion profiles are centrally peaked and then decrease with radius, as, for example, in the case of NGC1277 \citep{Trujillo+14}, which is the most extreme relic in F17, we expect to measure a smaller dispersion since we cover a larger spatial aperture for the galaxies. In fact, looking at Figure~2 in F17, one can see that for r$\sim0.5-1$\Reff and beyond, the stellar velocity dispersion values are measured to be around $\sigma_{\star}\sim 250$\kms, thus perfectly consistent with the values we find for KiDS J0224-3143 and KiDS J0847+0112. KiDS J0847+0112, for which we infer a velocity dispersion $\sim50$ \kms lower, might still be consistent if it has a very peaked sigma profile similar to that of NGC1277, but it could also simply be a lower mass galaxy. 

%Although we do not derive full star formation histories for our objects, we can clearly see from the right panels of Figure~\ref{fig:ssp_plots} that the great majority of the SSP with non-zero weights used by \ppxf refers to models with very old age and super solar abundances. This is especially true for J0847+0112, where more than 95\% of the total weight is given to model older than 9 Gyrs. For KiDS J0224-3143, even though the final inferred mean mass-weighted age is very old, we observe a somehow more extended distribution and a small fraction of weights is also given to stars with younger ages. 
Finally, looking at the derived SFH, plotted in Figure~\ref{fig:sfh}, we seem to observe a `degree of relicness', which is in line with what was already reported by F17. The SFH of KiDS J0847+0112 is, in fact, more `extreme', especially when considering all the possible setups and uncertainties in the [$\alpha$/Fe] estimate. 

At the same time, in this sense, even KiDS J0314-3215 might be considered as a less extreme relic, since it is nevertheless dominated by very old stars (80\% of the stellar mass rather than 95\%). This means that this galaxy might have experienced little interaction after the first mass assembly episode, but this 'accreted' component comprises at most 20\% of its stellar mass.\footnote{We note that some papers still classify galaxies as relics when the recent star formation is up to 30\% in mass (e.g. \citealt{Quilis_Trujillo13}).} % \LEt{unclear. Please review} 
Further investigation is required to understand the implication of this on the size growth and time evolution of this object. 

In conclusion,  as in F17, we cannot make any secure statement on the basis of only a few objects (five in total). 
%nor it is our purpose to give a strict definition of `relicness' at this stage. 
We therefore stress the need for a larger galaxy sample, 
which is one of the main goals of the \INSPIRE\, survey.  %The existence of a "degree of relicness", already reported by F17 seem to be confirmed also for these two new objects
Hence,  we will further investigate the existence of the  `degree of relicness', especially in connection with the environment and other structure parameters (e.g. size or colours) and assess, in this way, the path that red nuggets undergo to become normal ETGs. 
%\chiara{BOTH J0224 and J0847 ARE IN FIELD. J0314 IN CLUSTER!!!}

\section{Conclusions and future perspectives}
\label{sec:conclusions}
Relic galaxies, the untouched survivors of the massive and compact high-z red nuggets,  provide a unique opportunity to track the formation of the in-situ component formed at high-z during the first phase of the formation scenario, which is mixed with the accreted one in normal massive ETGs.  
Since studying in detail the stellar population of a statistically large sample of high-z red nuggets would require very long integration times, relics are ideal laboratories that allow us to investigate the physical processes that shaped the mass assembly of massive galaxies in the high-z Universe. 
Hence, the {\sl \emph{conditio-sine-qua-non}} to build a complete and comprehensive model for the star formation and cosmic evolution of massive galaxies in the Universe is enlarging the sample of relic galaxies for which stellar population properties, including the IMF, can be studied in great detail. 
This is one of the main goals of the \INSPIRE \  project, which we introduced in this paper. 

Further to introducing the new survey, as well as its current status and the observing strategy, we presented here results based on three systems, representative of the whole sample. KiDS J0224-3143 is the most massive UCMG in our dataset, KiDS J0314-3215 is one of the most compact objects, and KiDS J0847+0112 is one of the objects at the lowest redshift. 

In particular, in this paper we have: 
\begin{itemize}
    \item[i)] presented the combined final XSH UVB+VIS spectrum for each object, also correcting for the different resolutions of the two arms;\footnote{We have neglected the NIR arm, which will be presented in the forthcoming papers of the survey since this arm requires a much more precise treatment given the presence of very strong atmospheric telluric absorptions}
    \item[ii)] calculated the values of the integrated (from a five-pixel aperture) stellar velocity and velocity dispersion with the \ppxf\, code, also testing the impact of the fitting setup on the final results;
    \item[iii)] inferred the [$\alpha$/Fe] abundance by comparing line-index strengths of Mg and Fe lines in the galaxies with those of \citet{Vazdekis15} SSP models with a Kroupa IMF, a range of ages, metallicities, and [$\alpha$/Fe] abundances (interpolated to cover a finer grid); 
    \item[iv)] inferred mass-weighted mean age and [M/H], performing full-spectral fitting of the spectra in the $3500-6500$\AA\ wavelength range and presented results for the two possible extreme cases in terms of smoothness of the inferred SFH; 
    \item[v)] estimated uncertainties on the kinematics and stellar population parameters repeating the fit multiple times, and changing at each run the noise level, the degree of the polynomial used to adjust the continuum, and the good-pixel region; 
    \item[vi)] confirmed two of the three galaxies as relics, proving that they formed the majority ($>95\%$) of their mass at very high redshifts ($z\ge 6$), and,  in addition, they are also characterised by a super-solar metallicity and $\alpha$ abundance, similarly to the three confirmed local relics of \citet{Ferre-Mateu+17}. 
\end{itemize}

In conclusion, we confirmed two of the three galaxies analysed here as relics. If we extrapolate this result to the whole \INSPIRE\, sample, this means that we might confirm $\sim27$ UCMGs as relics and thus increase by a factor of $\sim 8$ the number of known relic galaxies, also extending the sample to relatively higher redshifts.

Once the full \INSPIRE\, sample is fully reduced and analysed, we will be able to compute precise number density of relics in the redshift window $0.1<z<0.5$ and compare our findings with the predictions from state-of-the-art hydrodynamical cosmological simulations. 
This will allow us to study, in detail, the mechanisms responsible for the dramatic size growth measured from $z\sim1$ to the present-day Universe and thus %put constrains and 
disentangle the possible merger scenarios. 
We will also investigate the formation scenario and mass assembly of these extremely compact and massive passive objects, for both relics and younger UCMGs.  
In particular, the presence of a conspicuous number of young very compact and very massive galaxies will seriously challenge current hydrodynamical simulations that predict the formation of such objects only in gas-rich environments of the young, high-redshift Universe \citep[e.g. ][]{ Zolotov15}. Thus, a measurement of the relative abundance of post-star-forming and old compact galaxies and its evolution with redshift will provide a key test for any models of massive galaxy formation.

%Furthermore, thanks to the wide wavelength range covered by X-Shooter, we will obtain, at least for the relics with the highest SNR spectra, a precise estimate of the low-mass end slope of the stellar IMF. This will help in understanding in detail the mechanisms with which stars formed in the high-z Universe and to possible reconcile the heavily debated results in this field. 
In addition, by inferring the stellar population parameters, including the IMF slope at the low-mass end (at least for the objects with highest S/N spectra) of relics, we will investigate the mechanisms that regulated the star formation at high-z, when the density and temperature of the Universe were much higher.  
According to theoretical work \citep[e.g. ][]{Chabrier_2014}, high density, temperature, and turbulence of the gas are key parameters that drive the fragmentation of molecular clouds. Higher density and a higher temperature make fragmentation easier, forming more dwarf stars, hence a bottom-heavier IMF. 
%that our team recently found in T18 and \citep[][hereafter S20]{Scognamiglio20} in the Kilo Degree Survey Data Release 3 (KiDS, \citealt{deJong+17_KiDS_DR3}),
%found in KiDS DR3 \citep{deJong+17_KiDS_DR3}. 
%It will be, for instance, interesting to compare the ratio between ordered and disordered motion ($V/\sigma$) in relics which permits to disentangle between a gas-rich driven merger triggering an intense burst of centralized star formation, from a rapid star formation at a time when the Universe was very dense. In fact, in the case of gas-rich merger, a higher rotation (disk-like structure), and thus a higher $V/\sigma$, is expected \citep{Wellons16}.

Finally, measuring the stellar velocity dispersion from spectroscopic data and stellar masses via stellar population analysis will allow us to compute the ratio between stellar and dynamical mass (M$_{\star}$/M$_{\mathrm{dyn}}$) for individual objects, a potentially valuable tracer of the likely mechanism by which galaxies grow \citep[e.g. ][]{Hopkins+09_DELGN_IV, Hilz+13}.  Specifically, if  galaxies experience  merger-driven growth, the ratio measured within the effective radius should decrease with time \citep{Tortora+14_DMevol, Tortora+18_KiDS_DMevol}; this decrease would be stronger in the case of minor mergers \citep{vandeSande13}. Hence, under the hypothesis that relics experienced very little or no mergers at all, they should all show a very high M$_{\star}$/M$_{\mathrm{dyn}}$, and thus a small amount of dark matter within the effective radius.
However, for one for the confirmed relics in the local Universe, Mrk~1216, \citet{Buote19} found the opposite from Chandra X-ray observations: the internal (within \Reff) DM fraction of this object is two times higher than the typical fractions found in nearby massive normal-sized ETGs. 

The only way to investigate further is to enlarge the sample of confirmed relics for which these kinds of studies can be performed, also pushing the redshift boundaries and finding systems living in different environments. This is indeed what we will achieve thanks to \INSPIRE. 

\medskip

\begin{acknowledgements}
The authors thank the referee for a very constructive report which helped improving the final quality of the manuscript.\\ CS is supported by a Hintze Fellowship at the Oxford Centre for Astrophysical Surveys, which is funded through generous support from the Hintze Family Charitable Foundation.  CS is also very grateful to Ortwin Gerhard and his `Dynamics Group' at the 
Max-Planck-Institut f\"{u}r Extraterrestrische Physik (MPE, Garching by Munich) for interesting and constructive discussions. \\
CT, AG, LH and SZ acknowledge funding from the INAF PRIN-SKA 2017 programme 1.05.01.88.04. \\
GD acknowledges support from CONICYT project Basal AFB-170002. \\
AFM has received financial support through the Postdoctoral Junior Leader Fellowship Programme from La Caixa Banking Foundation (LCF/BQ/LI18/11630007). \\
NRN acknowledges financial support from the One hundred top talent programme of Sun Yat-sen University, grant N. 71000-18841229. \\
DS is a member of the International Max Planck Research School (IMPRS) for Astronomy and Astrophysics at the Universities of Bonn and Cologne. 

\end{acknowledgements}

\bibliography{biblio_INSPIRE.bib} 

\begin{appendix}
\section{Testing the effect of changing the grade of the polynomials in the fits}
\label{app:polinom}
As highlighted in the main body of the paper, we performed two different \ppxf\, fits, one for the kinematics and one for the stellar populations, with different configurations of the program.  In particular, we used an additive Legendre polynomial to correct for the continuum shape when inferring the velocity and velocity dispersion, while we used a multiplicative polynomial when calculating age and metallicity. 
In this Appendix, we motivate  %\LEt{not sure what you mean by motivating a choice. You make the choice? You select?} 
the choice of the polynomial degree that we use in the main text in both cases, and we also demonstrate the robustness of the presented results against the use of different values, within a wide range of possibilities. 

\begin{figure}[h]
    \centering
    \vspace{0.3cm}
    \includegraphics[width=8.5cm]{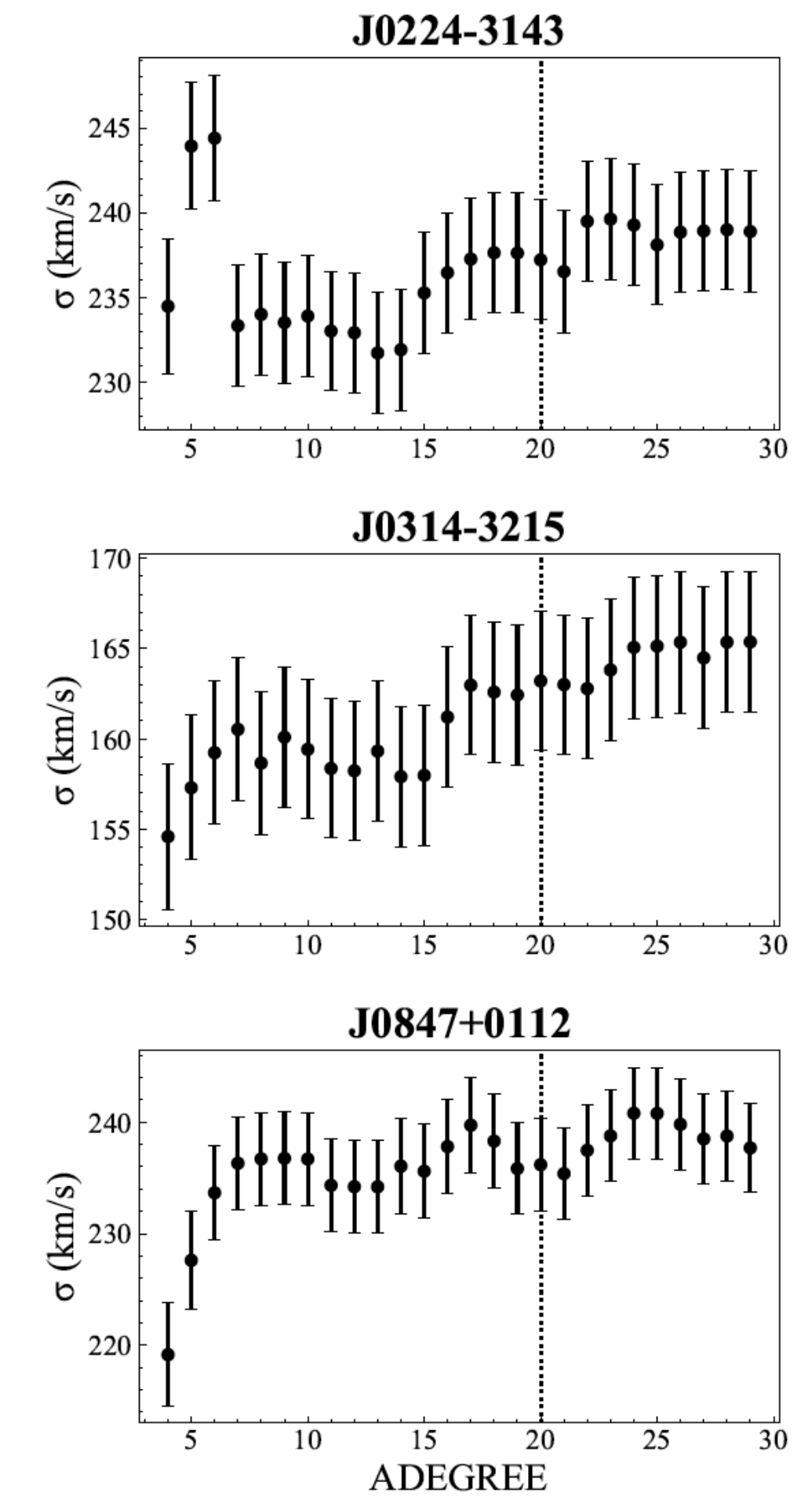}
    \caption{Velocity dispersion obtained as function of the additive polynomial degree in \ppxf.  We show here the results obtained using $\alpha$-enhanced SSP, and we refer the reader to Appendix~\ref{app:alpha} to see the effect that changing [$\alpha/$Fe] has on the resulting velocity dispersion values.  The results stabilise, for all the galaxies around ADEGREE$\sim20$, which is the final choice we adopted for the results given in Table~\ref{table:kinematics}. }
    \label{fig:kine_test}
\end{figure}

For the grade of the additive polynomial (ADEGREE), we test values from 4 to 29, with steps of 1 and finally chose ADEGREE = 20, which stabilises the results against changes of other parameters (e.g. stellar templates, good-pixel region, see the text for more details) while contemporarily minimising the reduced $\chi'^{2}$.
A demonstrative plot of the effect of the additive polynomial degree on the velocity dispersion is provided in Figure~\ref{fig:kine_test}.

\begin{figure}[h]
    \centering
    \includegraphics[width=8.5cm]{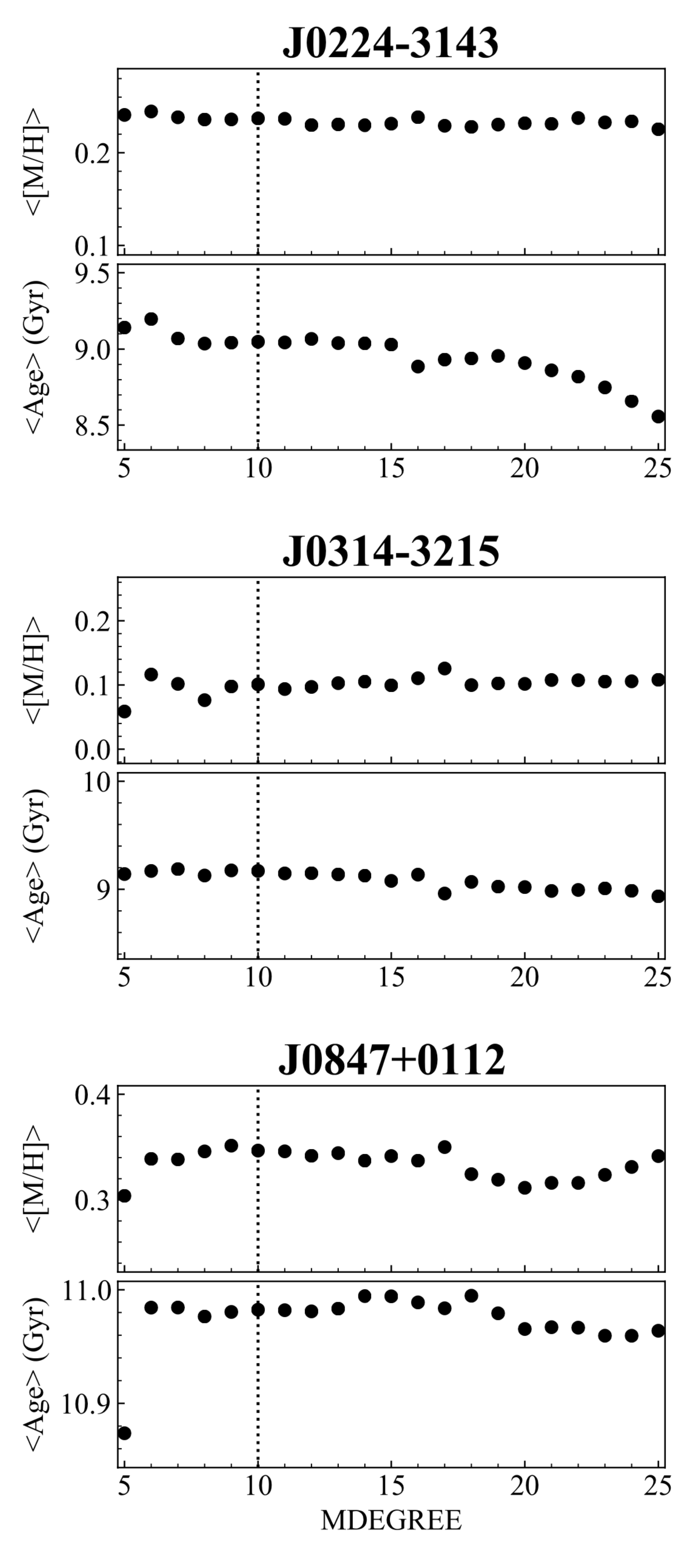}
    \caption{Mass-weighted mean metallicity (top) and age (bottom) as function of the multiplicative polynomial degree in \ppxf for each galaxy. The results stabilise for MDEGREE $\sim10$ (until $\sim20$), which is the final choice we adopted for the results given in Table~\ref{table:ssp}. }
    \label{fig:popu_test}
\end{figure}

For the multiplicative polynomial (MDEGREE) used to infer the stellar population parameters in Section~\ref{sec:population}, we tested degrees from 4 to 25 and finally chose MDEGREE = 10. The mass-weighted mean ages and metallicities derived by varying the grade of the multiplicative polynomial are shown in Figure~\ref{fig:popu_test}. For all the objects, the results are on average stable for 8<MDEGREE<20. Above this degree, the mean age of KiDS J0224-3143 becomes $\sim0.5$ Gyr younger, which is within the errors. We note that we report the results of this test for the REGUL = 0 case, but we conducted the same investigation for REGUL = MAX\_REGUL, finding fully consistent results. 

\section{Testing the effect of changing the [$\alpha$/Fe] values in the full-spectral fit}
\label{app:alpha}
\begin{figure}
\includegraphics[width=8.5cm]{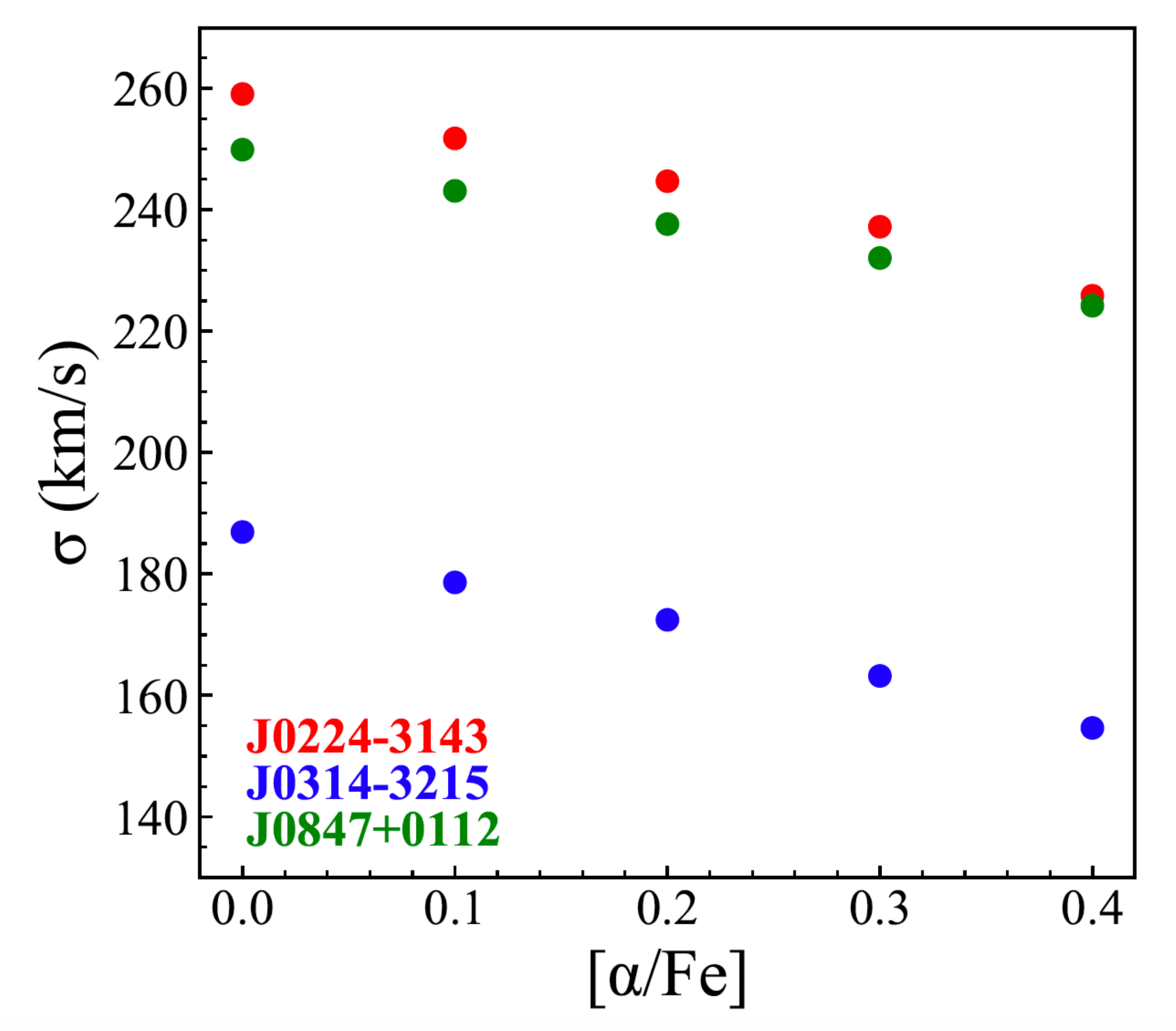}
\caption{Impact on stellar velocity dispersion of varying the [$\alpha$/Fe] abundances of the SSP models used for the \ppxf\, fit. Different colours indicate the results for the three different objects, as highlighted by the legend. }
\label{fig:varying_sigma}
\end{figure}

\begin{figure}
\includegraphics[width=8.5cm]{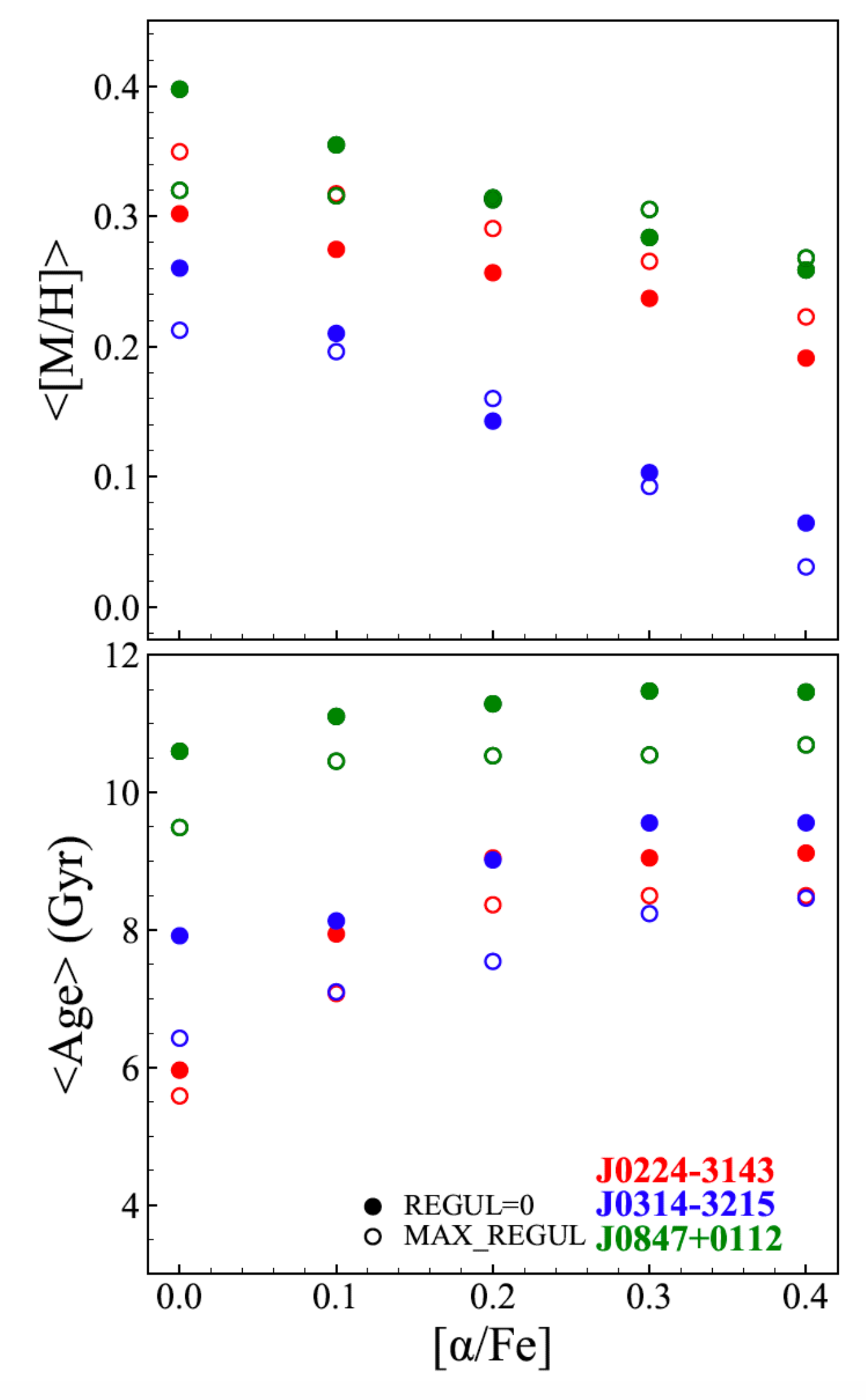}
\caption{Impact on stellar population parameters of varying the [$\alpha$/Fe] abundances of the SSP models used for the \ppxf\, fit. The top panel shows the variation on the inferred (mass-weighted) metallicity, whereas the bottom panel shows the variation in the (mass-weighted) stellar age. Filled symbols refer to the fits performed with REGUL = 0, while empty symbols refer to the regularised fits. Different colours indicate the results for the three different objects, as highlighted by the legend. }
\label{fig:varying_alpha}
\end{figure}

In the main body of this paper, we used two different approaches to derive the stellar population parameters. First, we inferred the [$\alpha$/Fe] from line-index measurements (see Fig.~\ref{fig:alphafe}), and then we obtained age and metallicity by performing a full spectral fitting with SSP models with the previously inferred $\alpha$ abundances, as templates. 
The line strength inference, although solid, might suffer from degeneracy between variation of [$\alpha$/Fe] and variation of single elemental abundances (e.g. Ti/Fe, which contaminates many optical iron lines). %, which is unfortunately not "tunable" from the SSP.   
Hence, here in this Appendix, we test the impact of using models with different  [$\alpha$/Fe] on the results presented in the paper. 
For this purpose, for each galaxy we repeated the \ppxf\, fit in 2D (varying age and metallicity) five times, each time using models with different $\alpha$ abundances (from 0.0 to 0.4 with a step of 0.1, see main body). We did that for both the REGUL = 0 and REGUL = MAX\_REGUL cases. The impact of a varying [$\alpha$/Fe] on the stellar velocity dispersion is shown in Figure~\ref{fig:varying_sigma} (the regularisation does not influence the $\sigma_{\star}$ values), while the impact on the stellar age and metallicity can be visualised in Figure~\ref{fig:varying_alpha}. Increasing [$\alpha$/Fe], the velocity dispersion and the metallicity decrease, while the inferred population gets older. However, this effect is much less evident for KiDS J0847+0112, the most extreme relic, and it is much larger for KiDS J0314-3215, which is the object with the most extended SFH. For KiDS J0224-3143, if one assumes a $[\alpha$/Fe]=0, then the inferred mass-weighted age becomes younger ($\sim 6$ Gyr), confirming what was already found in Figure~\ref{fig:sfh}. However, as already stated, this result would contradict the results inferred from the index-index plot (Fig.~\ref{fig:alphafe}).
This test confirms our result that KiDS J0847+0112 and KiDS J0224-3143 are indeed relic galaxies. 

\section{Testing the effect of different regularisation values on the stellar populations}
\label{app:regul}
\begin{figure*}
\begin{center}
\includegraphics[width=18.5cm]{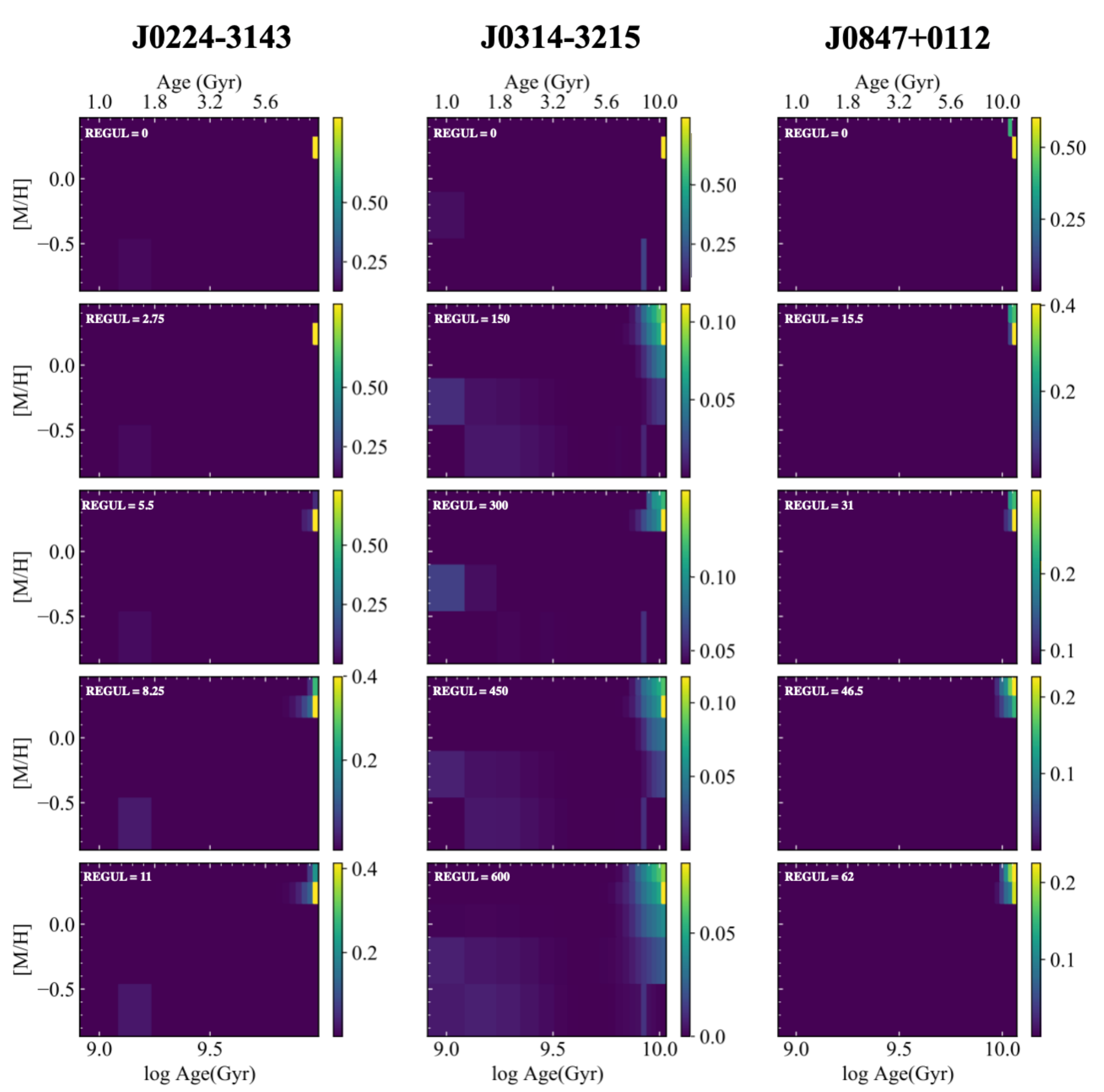}
\caption{2D density plot showing the weights attributed to the SSP models by \ppxf\, during the fit, in a $\log$ (age), [M/H] space (for simplicity, we also plot ages in linear scale on the top x-axis). Each column shows results for a different galaxy, while each row corresponds to the different results obtaining increasing the REGUL from 0 (top) to MAX\_REGUL (bottom). The weight fractions are reported on the vertical colour-bars, where yellow indicates the maximum weight and dark blue indicates zero weight. The upper limit of the age axis is slightly higher ($\sim10$\%) than the age of the Universe at the redshift of the objects, in order to take into account possible uncertainties in the value computed from the stellar population analysis.}
\label{fig:weightfract}
\end{center}
\end{figure*}

In order to give a meaningful interpretation of the template weights, in terms of the SFH and metallicity distributions of individual galaxies, a regularised fit should be tested. However, it is not trivial to properly set the REGUL parameter (see e.g. Sect.4.2 in \citealt{Shetty15}). For this reason, in this Appendix we undertake an investigation into the impact of different choices for the regularisation parameter on the results. 
In particular, following the prescriptions described in the \ppxf\, documentation, we: 
\begin{itemize}
    \item[{$\bullet$}] performed a number of unregularised fits (REGUL = 0), each time rescaling the input noise spectrum until obtaining a reduced $\chi'^{2}$ ($\chi^2$ divided by the number of fitted pixels, $N_{\mathrm{good\,pixels}}$) as close as possible to 1. All the other parameters for the fit were kept fixed (fitted region: $3500-6500$, ADGREE$=-1$); 
    \item[{$\bullet$}] performed a number of regularised fits, with the noise level set before, each time increasing the REGUL parameter until the new  $\chi^2$ increases by $\Delta \chi^2 = \sqrt{(2 \times N_{\mathrm{good\, pixels}}}$). In the remainder of this Appendix, we refer to the value of the REGUL keyword that optimises the $\chi'^{2}$ as MAX\_REGUL since, as already mentioned in Sect. \ref{sec:population}, it produces the smoothest possible star formation history. 
\end{itemize}
For each galaxy, we tested five different values of REGUL, between 0 and MAX\_REGUL (specifically $i\times$MAX\_REGUL, with i=[0.00, 0.25, 0.50, 0.75, 1.00]). As expected, when increasing the regularisation, the fit uses an increasing number of SSP templates, and the resulting solution, in terms of age and metallicity, is smoothed. However, this only causes very minor changes for the two newly confirmed relics, whereas it affects the results much more in terms of inferred mean ages and metallicities for KiDS J0314-3215. This can be visualised from Figure~\ref{fig:weightfract}, where we plot, for each of the galaxies (each column) in the form of a 2D density panel, the distribution of the weights (dark blue = 0, light yellow = maximum weight) that the \ppxf\, attributes to the SSP templates for the five cases with increasing regularisation (top row REGUL = 0, bottom row REGUL = MAX\_REGUL). 
No matter which regularisation value is used for KiDS J0847+0112, the fit uses only very old and metal-rich models. The mean values are almost identical for all cases. For KiDS J0224-3143, a young and metal-poor population  (age$\sim1.5$ Gyr) pops up when increasing the REGUL parameter. However, this contributes very little to the mass budget of the galaxy, as the inferred mean age is always consistently old.  
For KiDS J0314-3215, instead, increasing the smoothness of the fit, the distribution of the weights gets much more broad, and younger models with lower metallicities are used in the fit too. The inferred age is, in this case, roughly 1.5 Gyr younger. 

\end{appendix}

\end{document}